\PassOptionsToPackage{pdfpagelabels=false}{hyperref} 
\documentclass[twocolumn, trackchanges]{aastex63}

\usepackage{amsmath}

\usepackage{natbib}
\usepackage{textcomp}
\usepackage{gensymb}
\usepackage{graphicx}
\usepackage{appendix}

\usepackage{rotating}

\graphicspath{{./}{figures/}}

\submitjournal{ApJ}
\accepted{2021/02/18}

\shorttitle{A first Chandra view of the cool core cluster A1668}
\shortauthors{Pasini et al.}

\begin{document}

\title{A first Chandra view of the cool core cluster A1668: offset cooling and AGN feedback cycle}

\correspondingauthor{Thomas Pasini}
\email{thomas.pasini@hs.uni-hamburg.de}

\author[0000-0002-9711-5554]{T. Pasini}
\affiliation{Hamburger Sternwarte, Universität Hamburg, Gojenbergsweg 112, 21029 Hamburg, Germany}

\author[0000-0002-0843-3009]{M. Gitti}
\affiliation{Dipartimento di Fisica e Astronomia (DIFA), Universita` di Bologna, via Gobetti 93/2, 40129 Bologna, Italy}
\affiliation{Istituto Nazionale di Astrofisica (INAF) – Istituto di Radioastronomia (IRA), via Gobetti 101, I-40129 Bologna, Italy}

\author[0000-0001-9807-8479]{F. Brighenti}
\affiliation{Dipartimento di Fisica e Astronomia (DIFA), Universita` di Bologna, via Gobetti 93/2, 40129 Bologna, Italy}

\author[0000-0002-5671-6900]{E. O'Sullivan}
\affiliation{Harvard-Smithsonian Center for Astrophysics, 60 Garden Street, Cambridge, MA02138, USA}

\author[0000-0002-9112-0184]{F. Gastaldello}
\affiliation{INAF-IASF Milano, via E. Bassini 15, I-20133 Milano, Italy}

\author[0000-0002-8341-342X]{P. Temi}
\affiliation{Astrophysics Branch, NASA/Ames Research Center, MS 245-6, Moffett Field, CA 94035}

\author[0000-0003-1932-0162]{S. L. Hamer}
\affiliation{Department of Physics, University of Bath, Claverton Down, BA2 7AY, UK}

\begin{abstract}
We present a multi-wavelength analysis of the galaxy cluster A1668, performed by means of new EVLA and \textit{Chandra} observations and archival H$\alpha$ data. The radio images exhibit a small central source ($\sim$14 kpc at 1.4 GHz) with L$_{\text{1.4 GHz}}$ $\sim$6 $\cdot$ 10$^{23}$ W Hz$^{-1}$. The mean spectral index between 1.4 GHz and 5 GHz is $\sim$ -1, consistent with the usual indices found in BCGs. The cooling region extends for 40 kpc, with bolometric X-ray luminosity L$_{\text{cool}} = 1.9\pm 0.1 \cdot$ 10$^{43}$ erg s$^{-1}$. We detect an offset of $\sim$ 6 kpc between the cluster BCG and the X-ray peak, and another offset of $\sim$ 7.6 kpc between the H$\alpha$ and the X-ray peaks. We discuss possible causes for these offsets, which suggest that the coolest gas is not condensing directly from the lowest-entropy gas. In particular, we argue that the cool ICM was drawn out from the core by sloshing, whereas the H$\alpha$ filaments were pushed aside from the expanding radio galaxy lobes. We detect two putative X-ray cavities, spatially associated to the west radio lobe (cavity A) and to the east radio lobe (cavity B). The cavity power and age of the system are P$_{\text{cav}} \sim$ 9 $\times$10$^{42}$ erg s$^{-1}$ and t$_{\text{age}} \sim$5.2 Myr, respectively. Evaluating the position of A1668 in the cooling luminosity-cavity power parameter space, we find that the AGN energy injection is currently consistent within the scatter of the relationship, suggesting that offset cooling is likely not breaking the AGN feedback cycle.
\end{abstract}


\section{Introduction} \label{sec:intro}

In the last two decades, our understanding of the evolution of cool core galaxy clusters has led to a picture in which the cooling of the \textit{Intra-Cluster Medium} (ICM), the cold gas accreting onto the \textit{Brightest Cluster Galaxy} (BCG), and the feedback from the central radio source give birth to a tightly-connected cycle, known as Active Galactic Nuclei (AGN) feedback loop \citep[for reviews see e.g.][]{McNamara-Nulsen_2007, Gitti_2012, McNamara-Nulsen_2012, Fabian_2012}.
Multi-wavelength data provide strong evidences of this cycle: cavities in the ICM, revealed through deep X-ray observations and induced by the jets of the central radio galaxy \citep[e.g.,][]{McNamara_2000, Birzan_2004, Clarke_2004, Fabian_2006, Gentile_2007}, cold fronts \citep[e.g.,][]{Fabian_2006, Markevitch_2007, Gastaldello_2009, Ghizzardi_2010}, optical line emission \citep[][]{Crawford_1999, McDonald_2010, Hamer_2016} and dust filaments \citep[][]{VanDokkum_1995, Laine_2003} indicate an extremely complex and dynamical environment, whose physical processes are still to be completely understood. 
\\ \indent
Recently, a number of studies have revealed strong links between the central BCG, the X-ray core and the cluster dynamics \citep[][]{Sanderson_2009, Hudson_2010, Rossetti_2016}. In particular, spatial offsets between the BCG, the H$\alpha$ line emission and the X-ray emission peak \citep[e.g.][]{Haarsma_2010, Hamer_2012, Hamer_2016, Barbosa_2018} suggest that ICM sloshing and offset cooling, together with the AGN, can have a significant influence on the cluster evolution. Indeed, all these elements affect the activity of the central Supermassive Black Hole (SMBH) through motions of the gas, that could be able to regulate the cavity production and, consequently, the feedback cycle, since the ICM oscillates back and forth with respect to the central SMBH.
\\ \indent
This was recently discussed in \citet{Pasini_2019} for the cool core cluster A2495. Spatial offsets have been observed in this cluster, with the X-ray peak being separated by $\sim$ 6 kpc from the BCG and $\sim$ 4 kpc from the H$\alpha$ line emission peak. The analysis presented by the authors on two putative systems of X-ray cavities, hinted at in the shallow ($\sim 8$ ks) {\it Chandra} observation, suggests that even if cooling is not depositing gas onto the BCG core, the coupling between the AGN power output and the cooling rate is still consistent with the observed distribution for cluster samples. In a forthcoming publication we will present the detailed analysis of the deeper {\it Chandra} observations of A2495, recently allocated ($\sim 130$ ks, P.I. Gitti\footnote{Proposal Number 22800391}), which will be key to probe the presence of two pairs of ICM cavities and test the proposed scenario that the feeding-feedback cycle is not broken.
\\ \indent
A1668 was selected, along with A2495, from the ROSAT Brightest Cluster Sample (BCS; \citealt{Ebeling_1998}) by choosing objects with X-ray fluxes greater than 10$^{-11}$ erg cm$^{-2}$ s$^{-1}$ and, among these, by selecting those characterized by logL$_{\text{H}\alpha} >$ 40 from the catalogue of \citet{Crawford_1999}. Of the obtained sample of 13 objects, A2495 and A1668 still lacked \textit{Chandra} observations, that were obtained jointly with new VLA data (P.I. Gitti\footnote{Proposal Number 12800143}). \citet{Pasini_2019} have presented the results for A2495, making also use of H$\alpha$ line emission data and \textit{Hubble Space Telescope} (HST) archival images. In this work we combine the A1668 VLA and \textit{Chandra} new observations in order to study the interactions between the radio source hosted in the BCG and the ICM. As well as for A2495, we included H$\alpha$ line emission data from \citet{Hamer_2016}; on the other hand, no HST data are available for this cluster.
\\ \indent
A1668 was previously observed in the radio band by TGSS (TIFR GMRT Sky Survey), which gives an estimate for the 150 MHz flux density of 1589$\pm$ 159 mJy; \citet{Hogan_2015} performed a 5 GHz radio analysis (the data they used are not the same presented in this work), estimating a flux density of 21.0 $\pm$ 0.1 mJy. A1668 was recently included by \citet{Birzan_2020} in their sample of systems observed at 150 MHz by the LOw Frequency ARray (LOFAR, \citealt{vanHaarlem_2013}), showing the presence of large radio lobes, each extending for more than 50 kpc, and estimating a total flux density of 1.83 $\pm$ 0.44 Jy, consistent with TGSS.

Richness-based estimate of the mass provided values of M$_{200} \simeq$ 1.66$\cdot$10$^{14} \  \text{M}_{\odot}$ \citep{Andreon_2016} and M$_{2500} =$ 3.9 $\pm^{0.8}_{0.7} \cdot$ 10$^{13} \ \text{M}_{\odot}$ \citep{Pulido_2018}. The cluster's BCG, IC4130, shows a Star Formation Rate (SFR), estimated from extinction-corrected H$\alpha$ luminosity obtained from long-slit observations, of SFR = 2.5 $\pm$ 0.3 $\text{M}_{\odot}$ yr$^{-1}$ \citep[][]{Pulido_2018}, and extends for $\sim$ 85 kpc (diameter at the isophotal level of 25 mag/arcsec$^2$ in the B-band, \citealt{Makarov_2014}). \footnote{HyperLEDA catalog.} \citet{Edwards_2009} also presented IFU observations of the H$\alpha$ emission close to the BCG, finding a clear velocity gradient from positive values north of the centre to negative values at the south. They also argued that the line emitting gas is likely not at rest with respect to the BCG.
\\ \indent
In this work, we adopt a $\Lambda$CDM cosmology with H$_0$ = 73 km s$^{-1}$ Mpc$^{-1}$, $\Omega_M $= $ 1-\Omega_\Lambda $ = 0.3. The BCG redshift is $z$ = 0.06355 \citep[][]{Hamer_2016} and the luminosity distance is 273.7 Mpc, leading to a conversion of 1 arcsec = 1.173 kpc.

\section{Radio analysis}

\subsection{Observations and data reduction}

IC4130, the BCG of A1668, was observed with the EVLA on 2011 June 17th in the 1.4 GHz band, and on 2011 March 9th in the 5 GHz band, in A and B configurations respectively. Details of the observations are shown in Table \ref{tab:radiobs}.

\begin{table*}
	\centering 
	\begin{tabular}{c c c c c c}
		\hline
		\hline
		Frequency & Number of spw & Channels & Bandwith & Array & Total exposure time \\
		\hline
		5 GHz (C BAND) & 2 (4832 MHz - 4960 MHz) & 64 & 128 MHz & B & 3h59m21s \\
		1.4 GHz (L BAND) & 2 (1264 MHz - 1392 MHz) & 64 & 128 MHz &A & 2h59m28s \\
		\hline
	\end{tabular}
	\caption{Radio observations properties (project code SC0143, P.I. M. Gitti).} \label{tab:radiobs}
\end{table*}

The sources J1331+3030 (3C286) and J1327+2210 were used for both the observations as flux and phase calibrators, respectively.
The data reduction was performed using the NRAO Common Astronomy Software Applications package (CASA, version 5.3), applying the standard calibration procedure after carrying out an accurate editing of the visibilities with the CASA task {\ttfamily FLAGDATA}. We removed about 6$\%$ of the target visibilities at 5 GHz, whereas at 1.4 GHz the data were highly contaminated by Radio Frequence Interferences (RFI), thus producing a visibility loss of $\sim$ 40$\%$.
\\ \indent
We applied the standard imaging procedure, making use of the {\ttfamily CLEAN} task on a 7'' $\times$ 7'' region centered on the radio source. We took into account the sky curvature by setting the {\ttfamily gridmode=WIDEFIELD} parameter and used a two-terms approximation of the spectral model exploiting the MS-MFFS algorithm \citep{Rau_2011}.

\subsection{Results}
\label{cap:results}

We produced total intensity radio maps by setting {\ttfamily weighting = BRIGGS}, corresponding to {\ttfamily ROBUST 0}. This baseline weighting provides the best compromise between angular resolution (determined by long baselines) and sensitivity to extended emission (provided by short baselines). The uncertainty on the flux density measurements is 5$\%$, estimated from the amplitude calibration errors.

\begin{figure}
    \hspace{-0.6cm}
	\centering
	\includegraphics[height=30.5em, width=26.5em]{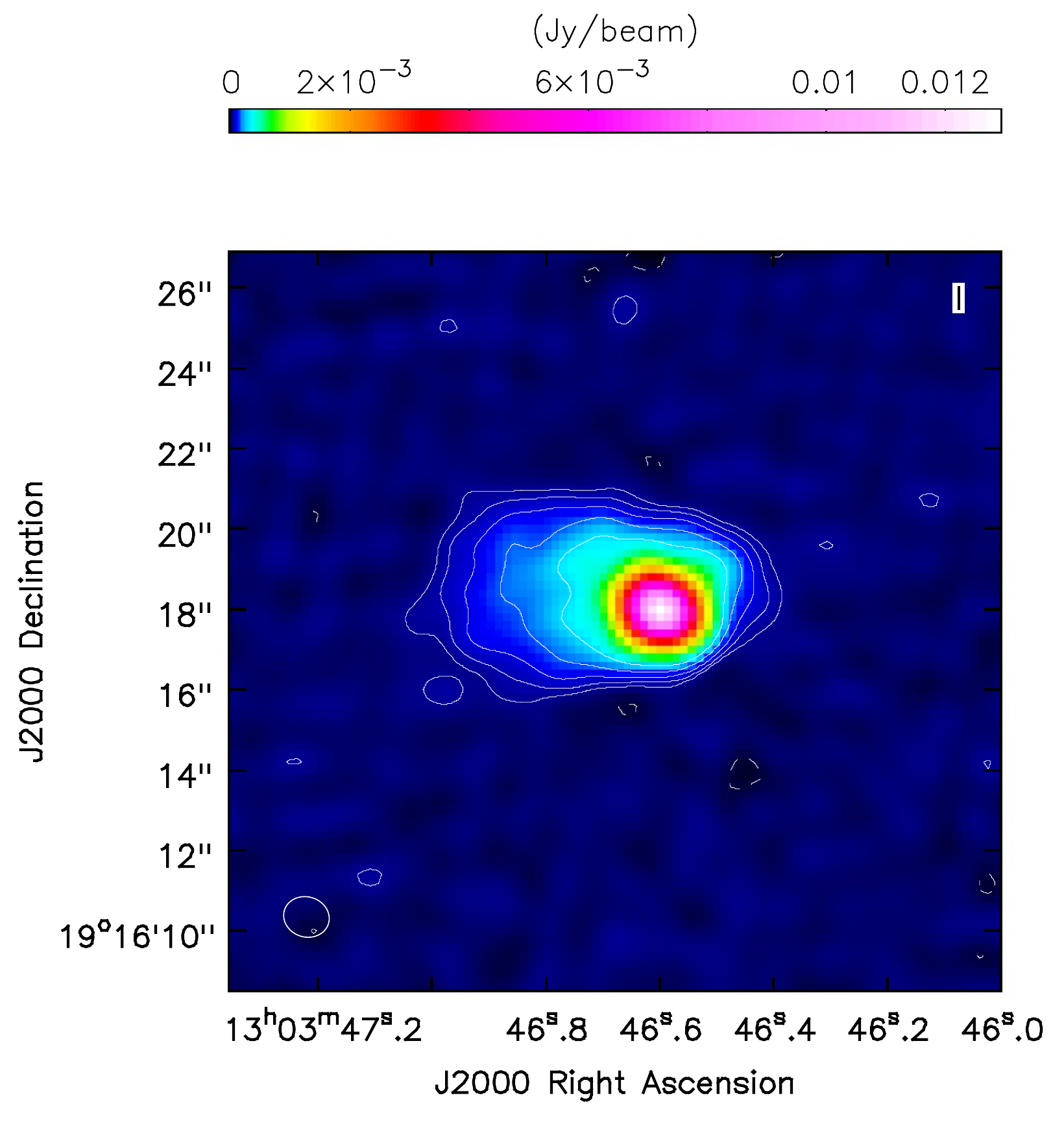}
	\caption{5 GHz VLA map ({\ttfamily ROBUST 0}) of the radio source hosted in IC4130, the BCG of A1668. The resolution is 1.14'' $\times$ 1.00'', with a rms noise of 6 $\mu$Jy beam$^{-1}$. Contours are at -3,3,6,12,24,48 $\cdot$ rms. The source flux density is 19.9 $\pm$ 1.0 mJy. The bottom-left white ellipse represents the beam.}
	\label{fig:Cbriggs}
\end{figure}

At 5 GHz (Fig. \ref{fig:Cbriggs}), the radio source exhibits a total flux density of 19.9 $\pm$ 1.0 mJy, consistent with \citet{Hogan_2015}, that corresponds to a luminosity of L$_{\text{5 GHz}} = (1.8 \pm 0.1) \cdot 10^{23}$ W Hz$^{-1}$. The rms noise is 6 $\mu$Jy beam$^{-1}$. The source stretches Eastwards for $\sim$ 11 kpc, with a minor axis of $\sim$ 5.7 kpc. There are no visible hints of larger emission up to the scale we are sensitive to (60 kpc with the VLA B configuration at 5 GHz). The equipartition magnetic field was estimated following the method described in \citet{Feretti-Giovannini_2008}, finding H$_{eq}$(5 GHz) = 8.7 $\pm$ 0.1 $\mu$G. 

\begin{table*}
	\centering 
	\begin{tabular}{c c c c c c c c}
		\hline
		\hline
		Band & Flux density & rms & beam &  Luminosity & Volume & Brightness Temperature & Equipartition Field \\ 
		&  [mJy] & [$\mu$Jy beam$^{-1}$] & [arcsec] & [10$^{22}$ W Hz$^{-1}$] & [kpc$^3$] & [K] & [$\mu$G] \\
		\hline
		5 GHz & 19.9 $\pm$ 1.0 & 6 & 1.14x1.00 &   16.8 $\pm$ 0.8 & 185 $\pm$ 22  & 39.6 $\pm$ 10.4 & 8.7 $\pm$ 0.1  \\
		1.4 GHz & 70.2 $\pm$ 3.5 & 17 & 1.44x1.08 & 59.1 $\pm$ 2.9  & 359 $\pm$ 30 & 1129.3 $\pm$ 241.2 & 10.3 $\pm$ 0.1  \\ 
		\hline
	\end{tabular}
	\caption{Radio properties of A1668 in the two bands observed. The axes of the radio galaxy are \textit{a} = 10.9 $\pm$ 1.3, \textit{b} = 5.7 $\pm$ 1.3 for the 5 GHz map and \textit{a} = 14.0 $\pm$ 1.3, \textit{b} = 7.1 $\pm$ 1.3 for the 1.4 GHz map. The flux density is estimated within 3$\sigma$ contours, while for the volume we assumed a prolate elissoid shape.} \label{tab:properties}
\end{table*}

\begin{figure}
	\hspace{-0.35cm}
	\centering
	\includegraphics[height=30.5em, width=26.3em]{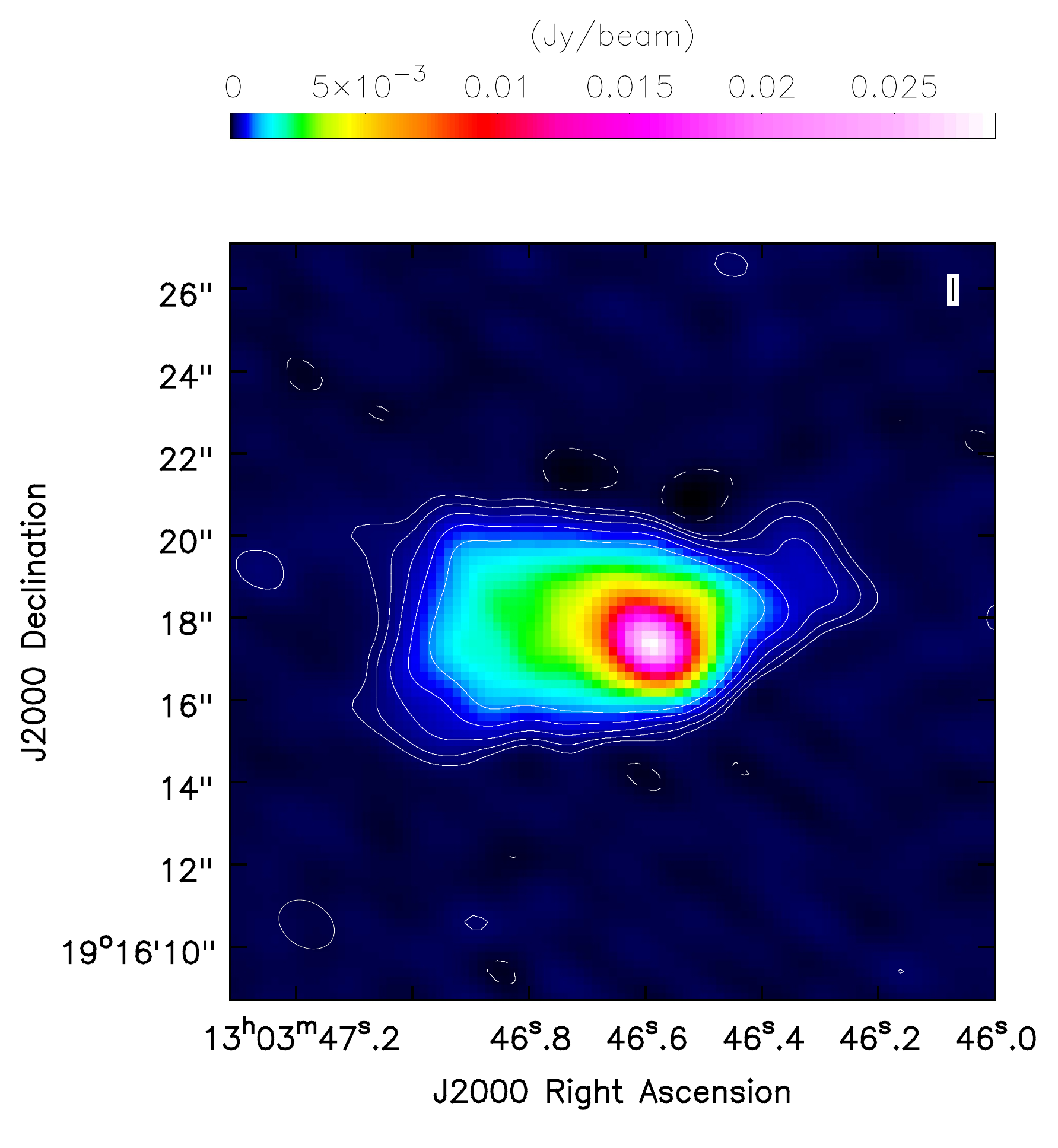}
	\caption{1.4 GHz VLA map ({\ttfamily ROBUST 0}) of the radio source hosted in IC4130. The resolution is 1.44'' $\times$ 1.08'', with a rms noise of 17 $\mu$Jy beam$^{-1}$. Contours are at -3,3,6,12,24,48 $\cdot$ rms. The source flux density is 70.2 $\pm$ 3.5 mJy. The bottom-left white ellipse represents the beam.}
	\label{fig:Lbriggs}
\end{figure}

The 1.4 GHz map ({\ttfamily ROBUST 0}, Fig. \ref{fig:Lbriggs}) shows no significant differences with respect to the 5 GHz emission. The source flux density is 70.2 $\pm$ 3.5 mJy and the rms is $\sim$ 17 $\mu$Jy beam$^{-1}$. The radio source scale is slightly larger ($\sim$ 14 kpc for the major axis, $\sim$ 7 kpc for the minor axis), with a more developed west lobe;  again, we did not detect any hint of larger scale emission up to our sensitivity scale (70 kpc with VLA A configuration at 1.4 GHz). Some cool core clusters show diffuse emission in the form of radio mini-halos \citep[e.g.,][]{Gitti_2004, Govoni_2009, Giacintucci_2014b}. \citet{Giacintucci_2017} define for mini-halos a minimum radius of 50 kpc since, at smaller radii, diffusion and other transport mechanisms are plausibly able to spread the relativistic electrons from the central AGN within their synchrotron radiative cooling time. In Fig. \ref{fig:Cbriggs} and Fig. \ref{fig:Lbriggs}, the radio emission is coincident with the optical BCG, and the small scale suggests that it can all be accounted to the AGN/radio galaxy. It is possible that diffuse emission larger than our sensitivity scale exists; however, given the extended double-lobe morphology of the LOFAR 150 MHz image presented in \citet{Birzan_2020}, the presence of a mini-halo in A1668 looks unlikely. The equipartition field is H$_{eq}$(1.4 GHz) = 10.3 $\pm$ 0.1 $\mu$G. Radio properties can be found in Table \ref{tab:properties}. 
\\ \indent
The radio source hosted in the centre of A1668 can be classified as a FRI galaxy, as demonstrated by both the morphology (asimmetric lobes, no hotspots) and the 1.4 GHz luminosity (L$_{\text{1.4 GHz}} = (6.3 \pm 0.3) \cdot 10^{23}$ W Hz$^{-1}$), that place IC4130 in the 70$^{\text{th}}$ percentile of the BCG radio luminosity function presented in \citet{Hogan_2015}.

\subsubsection{Spectral index map}

\begin{figure}
	\centering
	\includegraphics[width=25em]{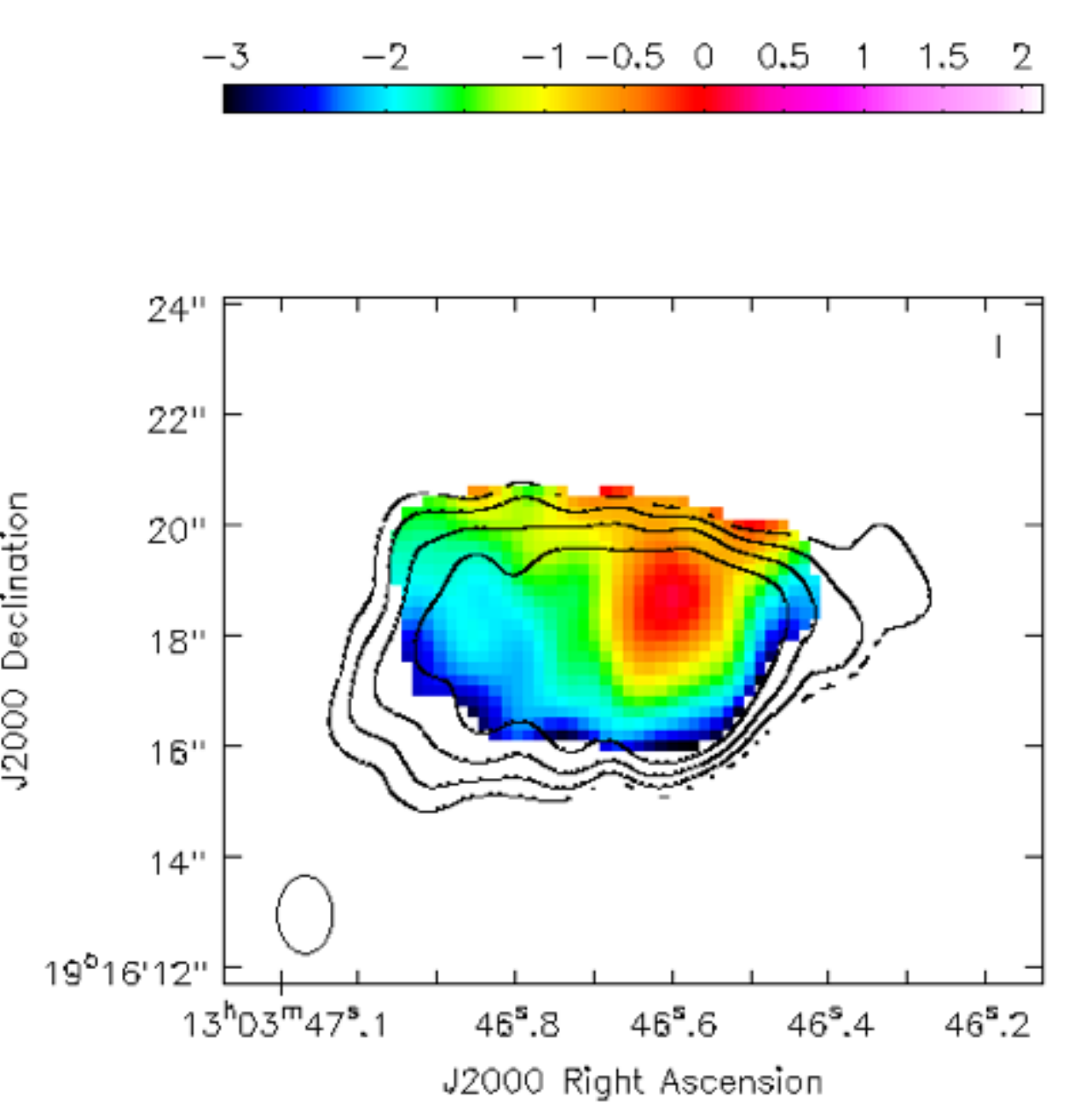}
	\caption{Spectral index map between 5 GHz and 1.4 GHz of the radio source hosted in IC4130. Contours are the same as Fig. \ref{fig:Lbriggs}, and typical errors range from $\Delta \alpha \simeq$ 0.1 for the inner and $\Delta \alpha \simeq$ 0.5 for the outer regions.}
	\label{fig:spix}
\end{figure}

The synchrotron spectrum follows a power law $S_{\nu} \propto \nu^{\alpha}$, where $\alpha$ is the spectral index. The spectral index map (Fig. \ref{fig:spix}) was generated using the CASA task {\ttfamily IMMATH}, combining 1.4 GHz and 5 GHz maps produced with matched {\ttfamily weighting=UNIFORM} (to enhance the resolution), {\ttfamily UVRANGE=6.5-152} , and a resolution of 1.4'' $\times$ 1.0''. The {\ttfamily UVRANGE} was set in order to be sensitive to the same baselines (thus, physical scales) for both observations.

Table \ref{tab:spindex} lists the peak, the extended and the total radio emission flux densities at 5 and 1.4 GHz, together with the estimated spectral index between the two frequencies.
The radio core exhibits a flat index ($\alpha \simeq 0$), as expected from optical thick regions where the radiation is self-absorbed. Moving towards the outskirt the spectrum becomes steeper, reaching $\alpha \simeq -2.5$ in the outermost part. The mean index is -0.99$\pm$ 0.06, consistent with the typical values found in BCGs \citep{Hogan_2015}. Table \ref{tab:spindex} summarizes the spectral index properties.

\begin{table}
	\centering 
	\begin{tabular}{c c c c }
		\hline
		\hline
		Region & $S_C \pm \Delta S_C$ & $S_L \pm \Delta S_L$ & $\alpha$ $\pm$ $\Delta \alpha$\\
		& [mJy] & [mJy] & \\
		\hline
		Peak & 7.5 $\pm$ 0.4 & 17.6 $\pm$ 0.9 & -0.67 $\pm$ 0.06\\
		Extended & 12.4$\pm$ 0.6 & 52.6 $\pm$ 2.6 & -1.13 $\pm$0.05\\
		Total & 19.9 $\pm$ 1.0 & 70.2 $\pm$ 3.5 & -0.99 $\pm$ 0.06\\
		\hline
	\end{tabular}
	\caption{The first column shows the flux density values at 5 GHz (C band), while the second displays the 1.4 GHz (L band) values. The third column presents the corrispondent spectral index values. The extended flux density was estimated as the difference between the total and the peak fluxes.} \label{tab:spindex}
\end{table}

\section{X-ray Analysis}

\subsection{Observation and data reduction}

A1668 was observed with the \textit{Chandra Advanced CCD Imaging Spectrometer} (ACIS), with the focal point on the S3 CCD, in cycle 12 (ObsID 12877, P.I. Gitti) for a total exposure of $\sim$ 10 ks. 
Data were reprocessed with CIAO 4.9 \citep{Fruscione_2006} using CALDB 4.2.1. We ran the {\ttfamily Chandra\_repro} script to perform the standard calibration process. After background flare removal, we used the {\ttfamily Blanksky} template files, filtered and normalized to the count rate of the source in the hard X-ray band (9-12 keV), in order to subtract the background. The final exposure time is 9979 s, with roughly $\sim$ 6800 net counts in a 100$"$ ($\sim$ 120 kpc radius region (0.5-2 keV) centered on the cluster.
\\ \indent
Point sources were identified and removed using the CIAO task {\ttfamily WAVDETECT}. Making use of optical catalogues, we found that no astrometry correction was necessary. Unless otherwise stated, the reported errors are at 68 $\%$ confidence level (1 $\sigma$).

\subsection{Results}

\subsubsection{Surface Brightness Profile}

\begin{figure}
	\centering
	\includegraphics[height=26.5em, width=25.5em]{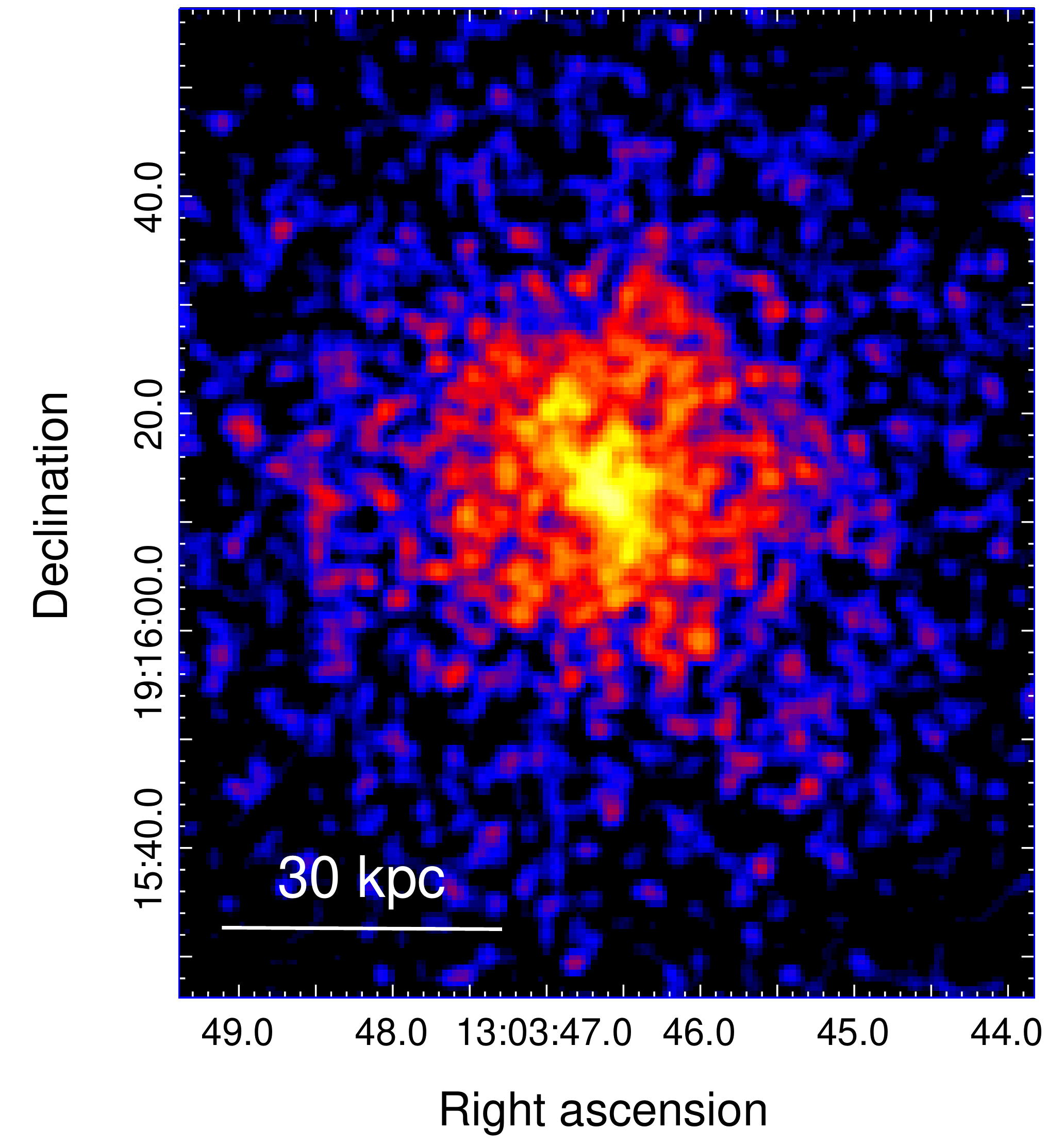}
	\caption{Chandra image of A1668 in the 0.5-2 keV band, smoothed with a gaussian filter with a 3 pixel radius.}
	\label{fig:a1668}
\end{figure}

\begin{figure}
	\centering
	\includegraphics[height=23em, width=25em]{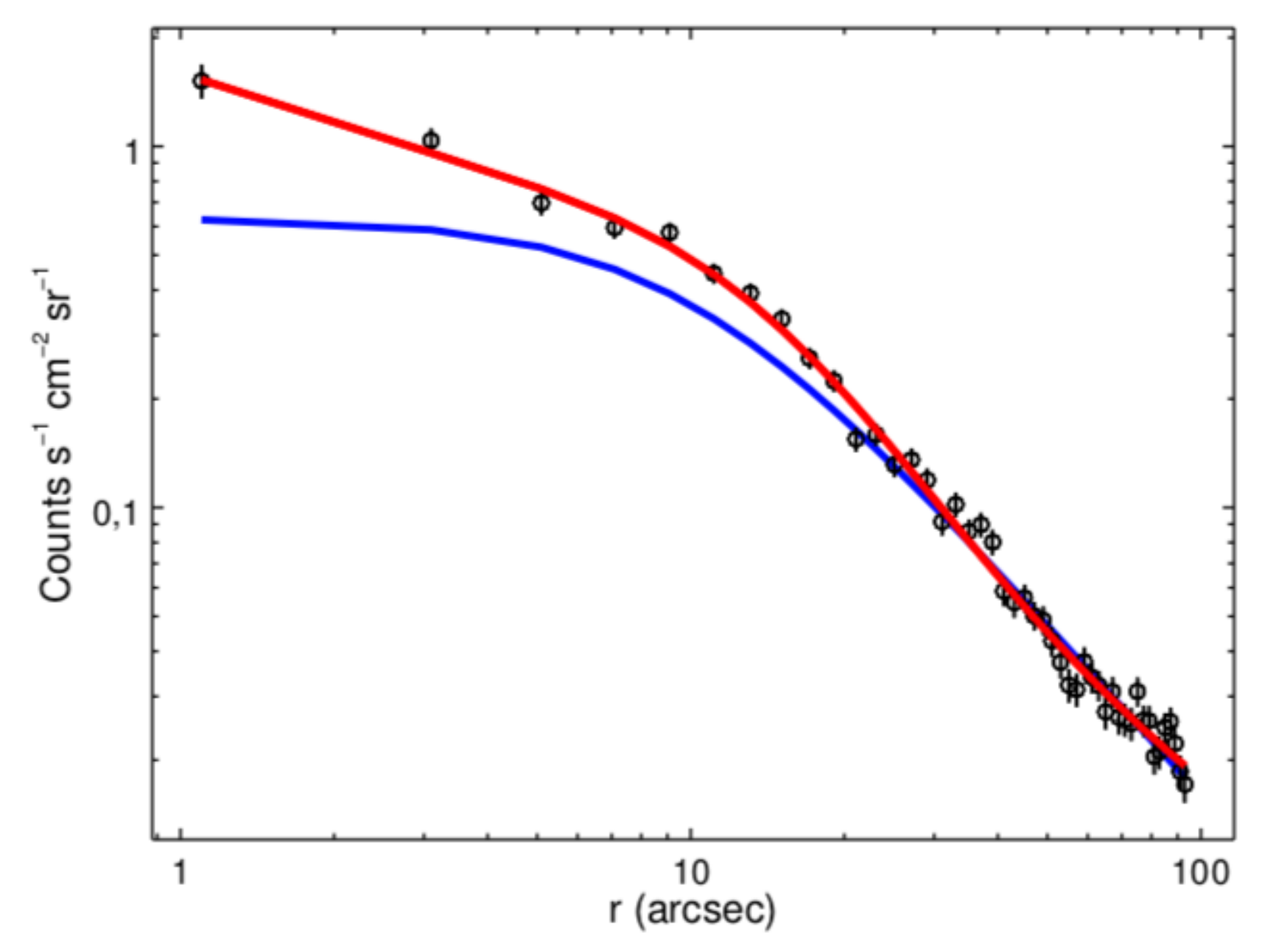}
	\caption{0.5-2 keV radial surface brightness profile of A1668. The blue line represents the single $\beta$-Model fit performed in the external 30"-100" interval and extrapolated to the center, while the red line is the double $\beta$-Model fit performed on every radius.}
	\label{fig:surbri}
\end{figure}

In Fig. \ref{fig:a1668} we show the smoothed 0.5-2 keV image of A1668.
The ICM exhibits a roughly circular and regular morphology on large scales ($>$ 30$"$ $\sim$ 35 kpc), while the cluster core shows a region with enhanced emission in the NE-SW direction.
Using the tool {\ttfamily SHERPA} \citep{Freeman_2001}, a surface brightness profile was produced from a background-subtracted, exposure-corrected image, making use of 2"-width concentric annuli centered on the X-ray peak. The profile was then fitted with a single $\beta$-Model \citep[][]{Cavaliere-Fusco_1976} over the external 30"-100" (35-120 kpc) interval, in order to exclude the whole core region\footnote{The assumption of 30$"$, that was already justifiable through visual inspection, will be furtherly supported, in Sec. \ref{sec:xanalisis}, by the estimate of the cooling radius}. The result of the fit ($\chi^2$/DoF $\sim$ 1.71) and its extrapolation to the core region is represented with the blue line in Fig \ref{fig:surbri}.
The best-fit values are: core radius {\ttfamily r0}=10.0 $\pm^{0.7}_{0.4}$ arcsec ($\sim$ 11.8 kpc), {\ttfamily beta}=0.43 $\pm ^{0.04}_{0.02}$ and central surface brightness {\ttfamily ampl}=0.64 $\pm^{0.10}_{0.04}$ counts s$^{-1}$ cm$^{-2}$ sr$^{-1}$.

The central brightness excess with respect to the $\beta$-Model is a strong indication of the presence of a cool core in A1668, as we expected from the selection criteria described in Sec. \ref{sec:intro}. This will also be confirmed by the spectral analysis (see Sec \ref{sec:xanalisis}). We therefore fitted the same profile on the entire radial range with a double $\beta$-Model \citep{Mohr_1999, LaRoque_2006},  represented with the red line in Fig. \ref{fig:surbri} ($\chi^2$/DoF $\sim$ 1.49), which provides a better description of the real trend; we found {\ttfamily r0$_1$}=15.6 $\pm^{4.2}_{2.7}$ arcsec ($\sim$ 18.3 kpc), {\ttfamily beta$_1$}=0.67 $\pm^{0.14}_{0.08}$ and {\ttfamily ampl$_1$}= 0.59 $\pm^{0.11}_{0.14}$ counts s$^{-1}$ cm$^{-2}$ sr$^{-1}$ for the first and {\ttfamily r0$_2$}=0.64 $\pm^{0.32}_{0.76}$ arcsec, {\ttfamily beta$_2$}=0.42 $\pm^{0.02}_{0.01}$ and {\ttfamily ampl$_2$}=1.78 $\pm^{0.75}_{3.07}$ counts s$^{-1}$ cm$^{-2}$ sr$^{-1}$ for the second $\beta$-Model.

\subsubsection{Spectral Analysis}
\label{sec:xanalisis}

\begin{figure}
	\hspace{-0.5cm}
	\includegraphics[height=30em, width=29em]{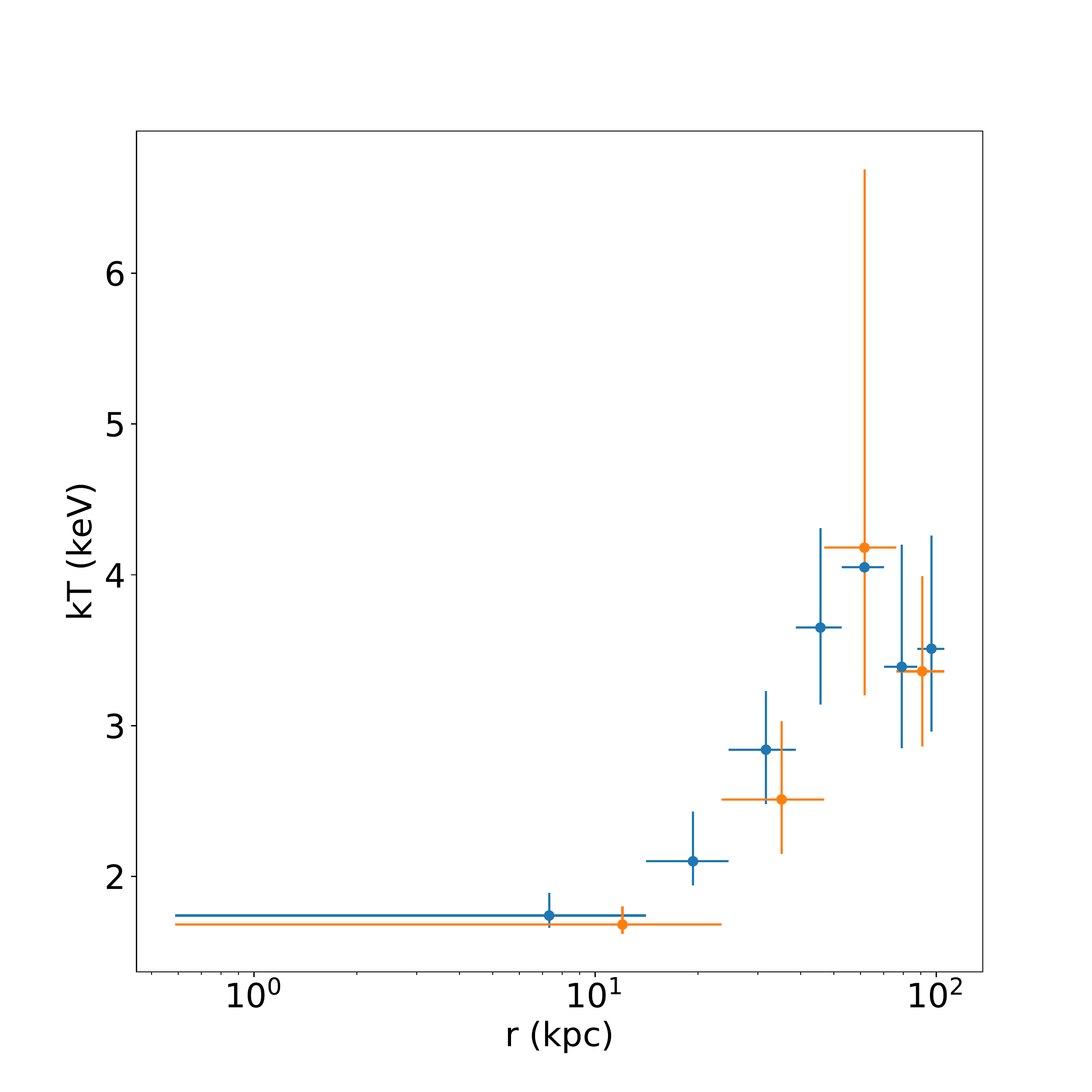}
	\caption{Projected (blue) and deprojected (yellow) temperature profile of A1668. Bars in the x-axis represent the range of the extraction rings, while in the y-axis are the errors for the temperature values.}
	\label{fig:temp}
\end{figure}

Spectra were extracted with the CIAO task {\ttfamily specextract} in the 0.5-7 keV band; the extraction was made from a series of concentric rings centered on the X-ray peak. Each region contains at least $\sim$ 1000 net counts. Background spectra were also extracted from the {\ttfamily Blanksky} files of each region. We individually fitted every spectrum via {\ttfamily Xspec} \citep[][vv.12.9.1]{Arnaud_1996} using a {\ttfamily phabs*apec} model, approximating an absorbed,  collisionally-ionized diffuse gas. The redshift was fixed at z=0.06355 and the hydrogen column density was fixed at N$_H$ = 2.20 $\cdot$ 10$^{20}$ cm$^{-2}$ (estimated from \citealt{Kalberla_2005}). The normalization parameter and the temperature kT were left free to vary. The observation was too shallow to allow us to fit metallicity, which was instead kept fixed at a value of 0.3 Z$_\odot$\footnote{This value was assumed after we tried to leave the metallicity free to vary. However, errorbars were too large to keep it thawed.}$^,$\footnote{The exploited abundance table is from \citet{Grevesse-Anders_1989}}. Note that these fits do not take in to account projection effects. The best-fitting parameters are listed in Table \ref{tab:proje}. The projected temperature profile of A1668 is shown in blue in Fig \ref{fig:temp}.

\begin{table}[!htb]
	\footnotesize
	\centering 
	\begin{tabular}{c c c c c}
		\hline
		\hline
		$r_{\text{min}}$-$r_{\text{max}}$ & $r_{\text{min}}$-$r_{\text{max}}$ & Counts & kT $\pm \sigma_{\text{kT}}$ &  $\chi^2/$DoF \\
		
		[arcsec] & [kpc] & & [keV] & \\
		\hline
		0 - 12 & 0 - 14 & 1346 (99.6 \%) & 1.74 $\pm ^{0.15} _{0.08}$ & 45/42 \\
		12 - 21 & 14 - 25 & 1274 (99.1 \%) & 2.09 $\pm ^{0.33}_{0.16}$ & 62/47 \\
		21 - 33 & 25 - 35 & 1328 (97.8 \%) & 2.84 $\pm ^{0.39}_{0.36}$ & 68/52 \\
		33 - 45 & 35 -  53 & 1111 (96.9 \%) & 3.65 $\pm ^{0.66}_{0.51}$ & 60/50 \\
		45 - 60 & 53 - 70 & 1101 (93.8 \%) & 4.05 $\pm ^{0.90}_{0.66}$ & 78/58 \\
		60 - 75 & 70 - 88 & 947 (91.2 \%) & 3.39 $\pm ^{0.81}_{0.54}$ & 123/61 \\
		75 - 90 & 88 - 106 & 1006 (89.2 \%) & 3.51 $\pm ^{0.75}_{0.55}$ & 98/64 \\
		\hline
	\end{tabular}
	\caption{Fit results for the projected analysis. The first and second columns show the lower and upper limits of the extraction rings in arcsec and kpc, while the third column represents the number of source photons coming from each ring, with the percentage indicating their number compared to the total photons of the same region. In the last two columns we report the values of kT, with associated errors, and the $ \chi ^ 2 / $ DoF.} \label{tab:proje}
\end{table}

Projection effects were then taken into account extracting spectra from concentric rings centered on the X-ray peak, containing more than 1500 counts, and fitting them with a {\ttfamily projct*phabs*apec} model. Temperature and normalization were left free to vary, while column density, redshift and abundance were frozen at the same values of the projected analysis above. Results are listed in Table \ref{tab:depro}. The deprojected temperature profile of the cluster is shown in black in Fig. \ref{fig:temp}.

\begin{table*}[!htb]
	\footnotesize
	\centering 
	\begin{tabular}{c c c c c c c c c}
		\hline
		\hline
		$r_{\text{min}}$-$r_{\text{max}}$ & $r_{\text{min}}$-$r_{\text{max}}$ & Counts & kT & N(r) (10$^{-4}$) & Electronic Density & Pressure & Entropy & t$_{cool}$   \\
		
		[arcsec] & [kpc] & & [keV] & & [$10^{-2} \ $cm$^{-3}$] & [10$^{-11}$ dy cm$^{-2}$] & [keV cm$^2$] & [Gyr] \\
		\hline
		0 - 20 & 0 - 23.5 & 2506 (99.4 \%) & 1.68 $\pm ^{0.12} _{0.06}$ & 10.3 $\pm ^{0.6}_{0.6}$ & 2.67 $\pm ^{0.01}_{0.01} $ & 13.4 $\pm ^{1.1}_{0.4} $ & 19.5 $\pm ^{0.5}_{0.2}$ & 1.4 $\pm ^{0.1}_{0.1} $ \\
		20 - 40 & 23.5 - 46.9 & 2162(97.4 \%) & 2.51 $\pm ^{0.52}_{0.36}$ & 12.5 $\pm ^{0.8}_{0.8}$ & 1.07 $\pm ^{0.01}_{0.01} $ & 7.7 $\pm ^{1.4}_{1.0}$ & 51.8 $\pm ^{2.0}_{1.4}$ & 4.3 $\pm ^{0.6}_{0.6} $\\
		40 - 65 & 46.9 - 76.2 &  1862 (93.7 \%) & 4.18 $\pm ^{2.51}_{0.98}$ & 8.1 $\pm ^{0.8}_{0.8}$ & 0.44 $\pm ^{0.01}_{0.01} $ & 7.5 $\pm ^{5.1}_{2.4}$ & 219.9 $\pm ^{24.5}_{11.3}$ & 15.8 $\pm ^{5.2}_{5.1} $ \\
		65 - 90 & 76.2 - 105.5 & 1653 (89.1 \%) & 3.36 $\pm ^{0.63}_{0.50}$ & 16.9 $\pm ^{0.9}_{0.9}$ & 0.44 $\pm ^{0.01}_{0.01} $ & 4.2 $\pm ^{0.7}_{0.6}$ & 120.7 $\pm ^{3.5}_{3.0}$ & 11.7 $\pm ^{1.7}_{1.8} $ \\
		\hline
	\end{tabular}
	\caption{Fit results for the deprojected analysis. The first two columns report the limits of the annular regions and the number of source photons from each ring, with the percentage indicating their number compared to the total photons of the same region. The remaining columns report temperature, normalization factor, electronic density, pressure, entropy and cooling time. The fit gives $\chi^2/$DoF = 1.46.}
	\label{tab:depro}
\end{table*}

\begin{figure}
	\centering
	\includegraphics[height=28em, width=27.5em]{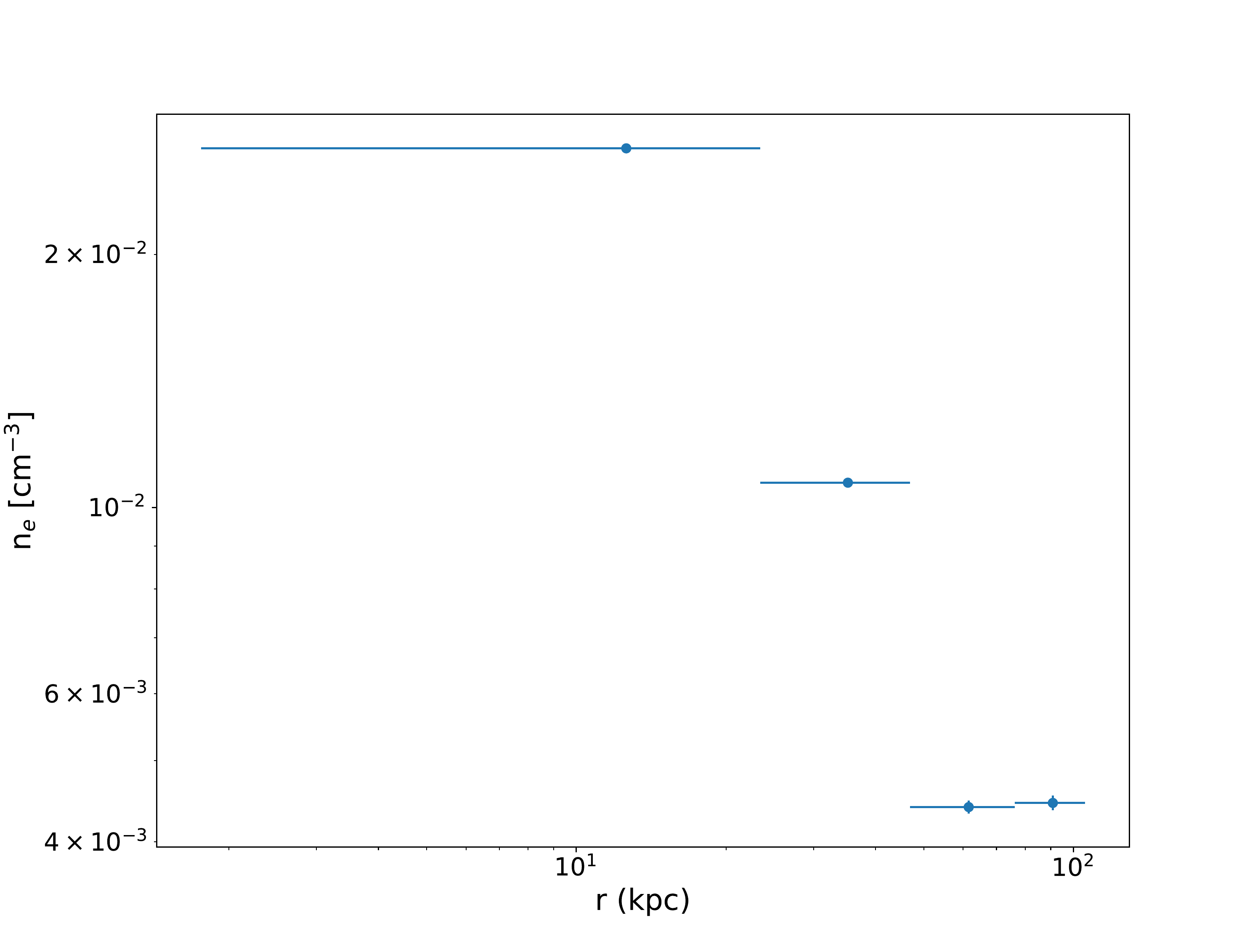}
	\caption{Density radial profile of A1668 derived from the deprojected analysis. Each bin defines an extraction region.}
	\label{fig:density}
\end{figure}

Following the same method described in \citet{Pasini_2019}\footnote{Note the typo in Eq. 4 of that paper}, we estimated the electronic density as :

\begin{equation}
\hspace{1cm}
n_e=\sqrt{10^{14} \bigg(\dfrac{4 \pi \cdot N(r) \cdot [D_A \cdot (1+z)]^2}{0.82 \cdot V}\bigg)}
\end{equation}

where N(r) is the {\ttfamily apec} normalization of the deprojected model, V is the shell volume and D$_A$ is the angular distance of the source, estimated as D$_A$ = D$_L$/(1+$z$)$^2$. Table \ref{tab:depro} lists the density values for each ring, with the results showed in Fig. \ref{fig:density}. 

\begin{figure}
	\centering
	\includegraphics[height=28em, width=27.5em]{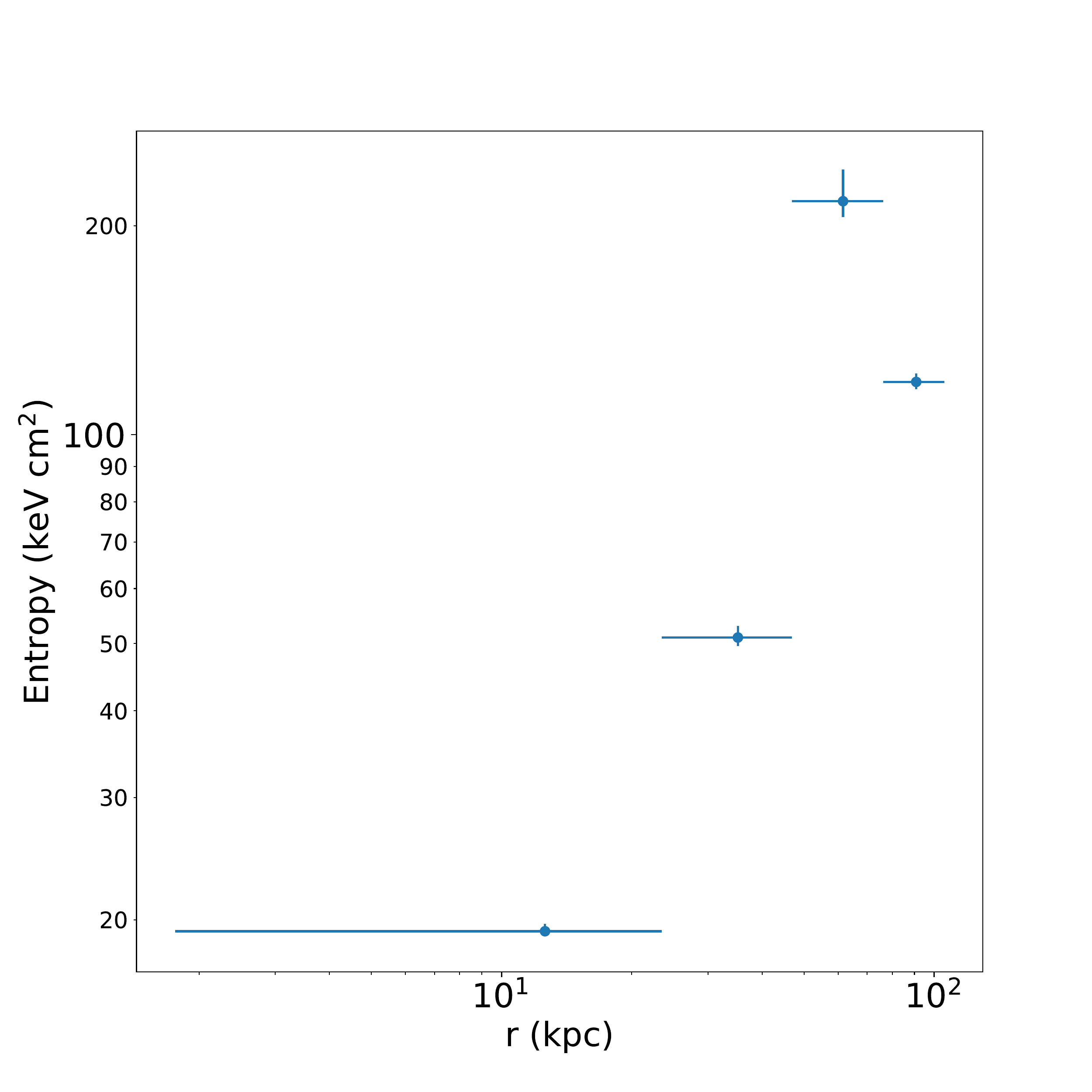}
	\caption{Entropy radial profile of A1668 derived from the deprojected analysis. Each bin defines an extraction region.}
	\label{fig:entropy}
\end{figure}

Making use of the deprojected temperature and density values, we can derive the cooling time, the pressure and the entropy for each bin. Table \ref{tab:depro} presents the pressure values, calculated as $p = 1.83n_e kT$, while the entropy, that was estimated as $S$ = $kTn_e^{-2/3}$, is presented in Fig. \ref{fig:entropy}.

\begin{figure}
	\centering
	\includegraphics[height=28.5em, width=27.5em]{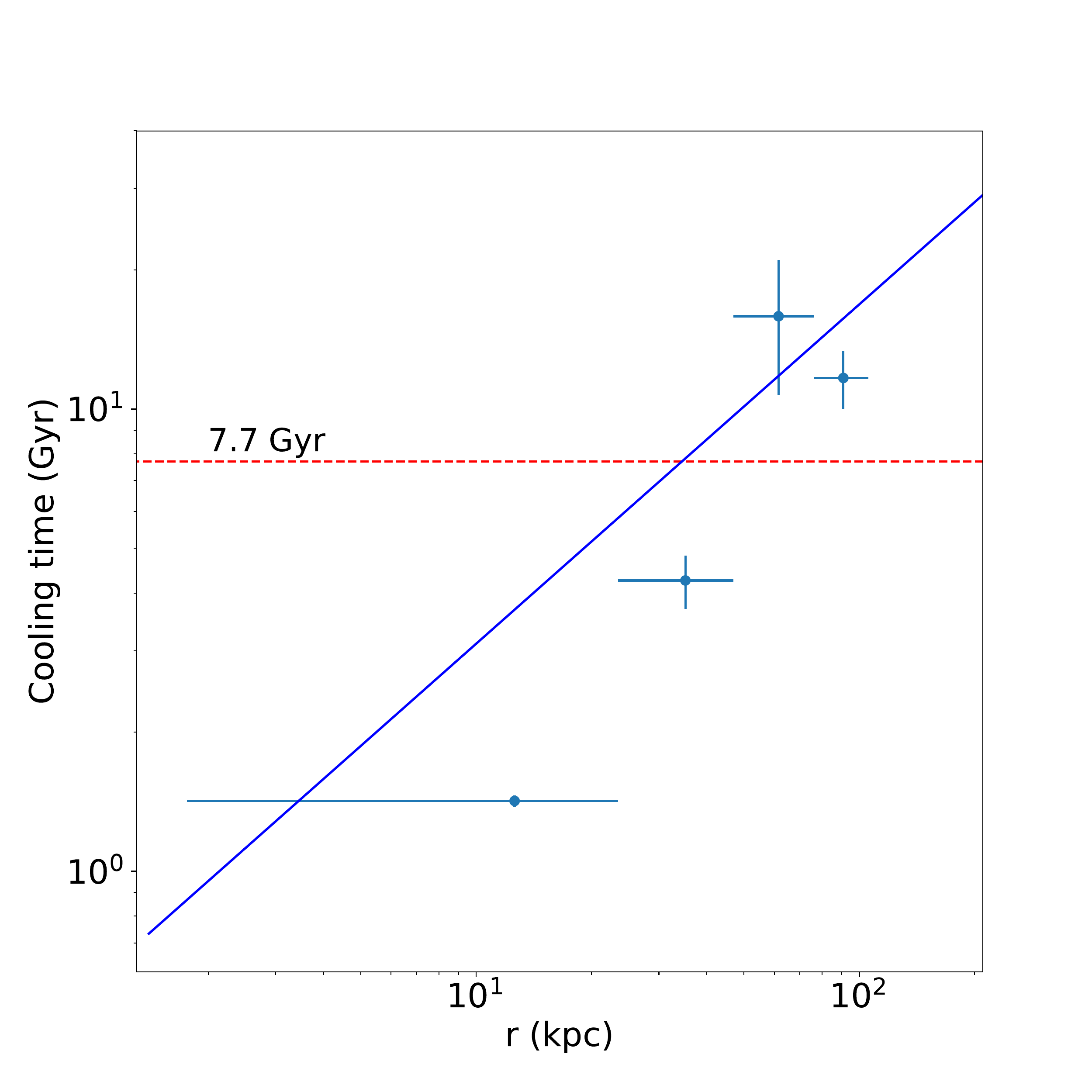}
	\caption{Cooling time profile of A1668. Each bin defines an extraction region. The blue line represents the best-fit function f(x)=(0.57$\pm$0.99)x$^{0.73 \pm 0.43}$, while the red line is t$_{\text{age}}$ = 7.7 Gyr.}
	\label{fig:tcool}
\end{figure}

The cooling time is defined as:

\begin{equation}
\hspace{1cm}
t_{\text{cool}}=\dfrac{H}{\Lambda (T)n_e n_p}=\dfrac {\gamma}{\gamma-1}\dfrac{kT(r)}{\mu X n_e(r) \Lambda(T)}
\end{equation}
\label{eq:tcool}

where $\gamma$=5/3 is the adiabatic index, $H$ is the enthalpy, $\mu \simeq$ 0.61 is the molecular weight for a fully ionized plasma, X $\simeq$ 0.71 is the hydrogen mass fraction and $\Lambda(T)$ is the cooling function \citep[][]{Sutherland-Dopita_1993}. Results are listed in Table \ref{tab:depro}, while the cooling time radial profile is shown in Fig. \ref{fig:tcool}.

We thus estimated the cooling radius of the cluster, i.e. the radius within which the ICM cooling is efficient, assuming t$_{\text{age}} \sim$ 7.7 Gyr, corresponding to the look-back time at z=1, as an upper limit for the cluster age.
Consequently, the intersection between the profile best fit (blue line) and t$_{\text{age}}$ (red line) defines the cooling radius of A1668, being $r_{\text{cool}} \approx 34  " \approx 40 \  \text{kpc}$.
\\ \indent
The bolometric X-ray luminosity emitted within this radius was estimated by extracting a spectrum from an annular region centered on the X-ray peak with r = $r_{\text{cool}}$. Projection effects were taken into account by using a second annular region with internal radius coincident with $r_{\text{cool}}$ and external radius $\sim$ 100$"$. By fitting both spectra with a {\ttfamily projct*phabs*apec} model, the bolometric luminosity inside the cooling region results $L_{\text{cool}}= 1.9 \pm 0.1 \cdot 10^{43} \ \text{erg \ s}^{-1}$.
Assuming a steady state cooling flow model, the \textit{Mass Deposition Rate} of the cooling flow of A1668 can be estimated as:

\begin{equation}
{\dot{\text{M}}} \simeq \dfrac{2}{5} \dfrac{\mu m_p}{kT} \cdot L_{\text{cool}}
\end{equation}

In this way, we obtain $\dot{\text{M}} \simeq 29.6 \pm 1.6 \ \text{M}_{\odot} \ \text{yr}^{-1}$.


As a different approach, we performed a further fit of the spectrum of the cooling region with a {\ttfamily phabs*(apec + mkcflow)} model, where the {\ttfamily apec} component approximates the ICM emission along the line of sight outside of the cooling region, while {\ttfamily mkcflow} is a multiphase component reproducing a cooling flow-like emission inside the cooling radius. As above, the abundance was fixed at 0.3 Z$_{\odot}$, while the temperature of the {\ttfamily apec} model was left free to vary and bounded to the {\ttfamily high temperature} parameter of {\ttfamily mkcflow}. Redshift and absorbing column density were fixed at the Galactic values (see above), while the {\ttfamily low temperature} parameter of {\ttfamily mkcflow} was fixed at the lowest possible value, $\sim$ 0.1 keV. The fit gives $\chi^2$/DoF = 105/100 and provides an upper limit of $\dot{\text{M}} < 5 \ \text{M}_{\odot} \ \text{yr}^{-1}$. The bolometric luminosity associated to the {\ttfamily mkcflow} model is L$_{\text{mkcflow}}= 3.2 \pm 0.1 \cdot 10^{41} \ \text{erg \ s}^{-1}$.
The difference between the two estimates of the mass deposition rate reflects the \textit{Cooling Flow (CF) problem}: observed mass deposition rates do not match expectations from the standard CF model, and heating contribution, likely produced by the central AGN, is required to balance the ICM radiative losses.

\section{Discussion}
\subsection{Radio-X-ray combined analysis}
\label{sec:radiox}

\begin{figure*}
	\includegraphics[height=32em, width=28.5em]{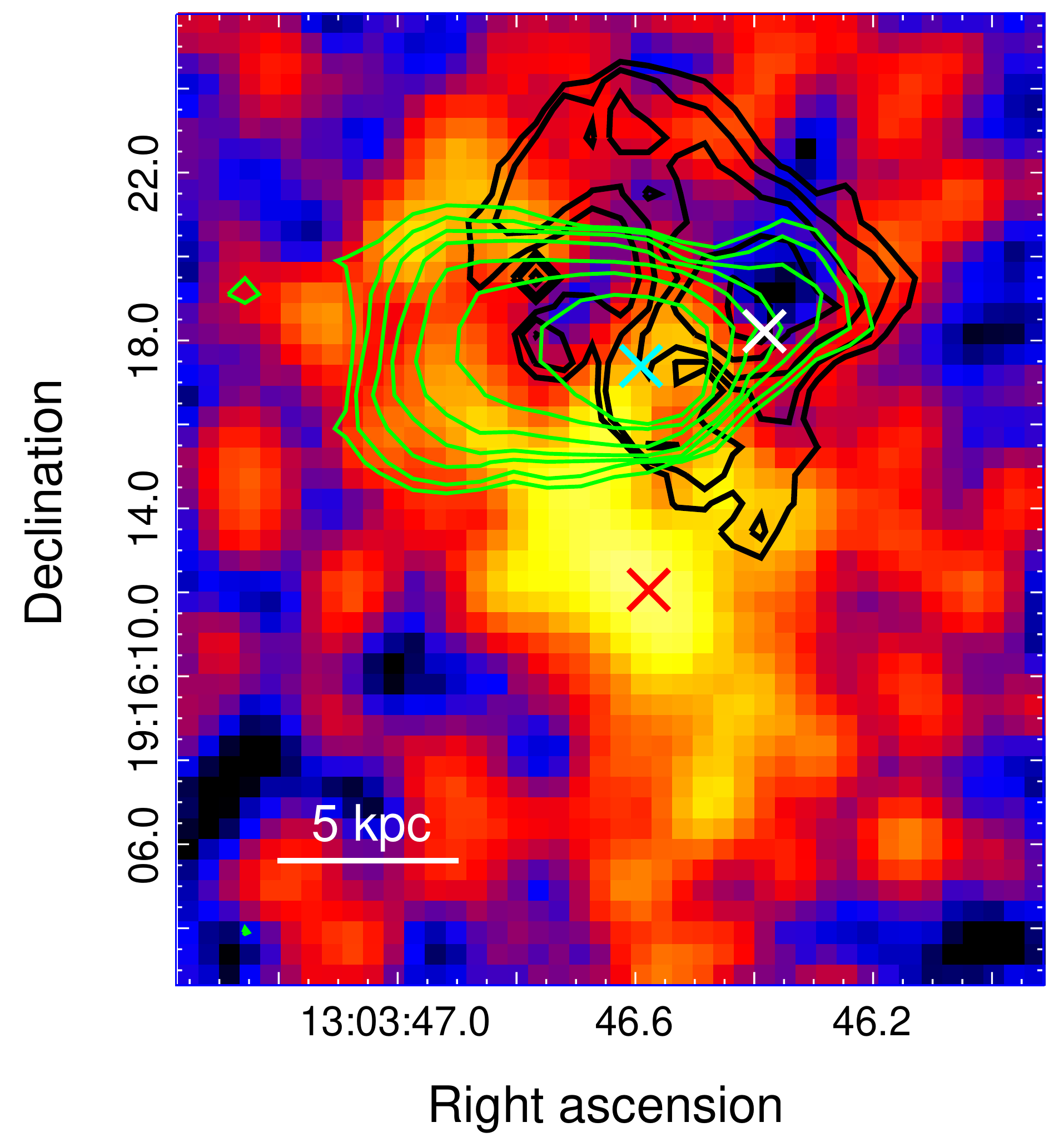}
	\includegraphics[height=31.8em, width=26.5em]{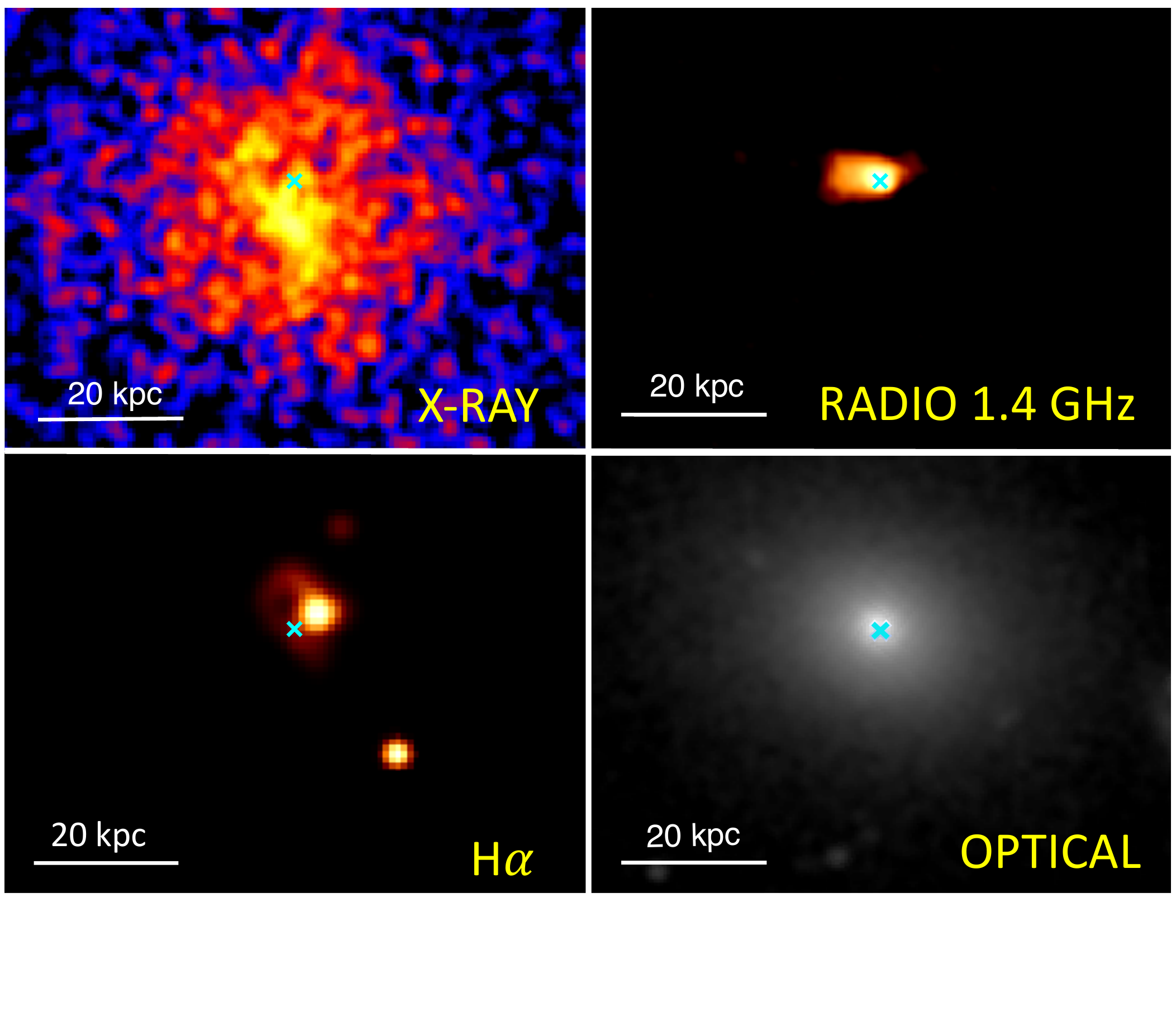}
	\caption{\textit{Left Panel}: 1.4 GHz radio (green) and H$\alpha$ (black) contours overlaid on the 0.5-2 keV X-ray image, zoomed towards the cluster centre. The cyan cross represents the X-ray emission centroid, coincident with the BCG centre; the red and white crosses are the X-ray and H$\alpha$ peaks, respectively. \textit{Right Panel}: From left to right, from top to bottom: 0.5-2 keV, 1.4 GHz, H$\alpha$ and optical (SDSS) images of A1668. All images are centered on the X-ray peak. The cyan cross represents the BCG.}
	\label{fig:multi}
\end{figure*}

In order to investigate the interactions between the cooling ICM and the BCG, we overlaid the 1.4 GHz radio contours on the X-ray 0.5-2 keV cluster image. Since we are interested in the core region, in Fig. \ref{fig:multi} we show the resulting image, zoomed in the central $30\times30$ kpc.

The cluster X-ray cool core exhibits an elliptical nuclear region. Exploiting optical catalogues, we found that the radio galaxy is coincident with the BCG nuclear region, as expected. The emission centroid of the large-scale X-ray emission ($RA$=13$^{\text{h}}$03$^{\text{m}}$43.6$^{\text{s}}$, $DEC$=+19$\degree$16$^{\text{m}}$17.4$^{\text{s}}$), defined as the center of the isophotes, lies within this region, too. On the other hand, the X-ray peak ($RA$=13$^{\text{h}}$03$^{\text{m}}$46.6$^{\text{s}}$, $DEC$=+19$\degree$16$^{\text{m}}$12.2$^{\text{s}}$), is found to the south of the nucleus of the BCG, exhibiting a significant offset of $\sim$ 5.2$"$, corresponding to $\sim$ 6 kpc. 
\\ \indent 
In order to check for the possible presence of a field point-source that could bias the detection of the X-ray peak, we extracted a spectrum from a $\sim$ 4$"$ circular region centered on the peak, and fitted it with two models: {\ttfamily phabs*apec} and {\ttfamily phabs*(apec+powerlaw)}. The first fit gave $\chi^2/\text{DoF} \sim$ 78/79, while for the second $\chi^2/\text{DoF} \sim$ 71/77; the F-stat method was then applied in order to check if the addition of the {\ttfamily powerlaw} component provided a significant improvement of the fit. We obtained an F-value of 3.4 and p=0.035, correspondent to a null hypothesis probability of 1-p=0.965. This suggests that the addition of the point-source emission component is not statistically significant. 
As a further check, we looked for possible point-sources in high energy, optical and infrared catalogues, as well as in a harder band (4-7 keV) X-ray image; however, we did not detect any point source coincident with the X-ray peak. We thus conclude that the peak detection is likely not biased, and therefore the offset is real.
\\ \indent
This is analogous with what was found in A2495, that presents a similar-scale offset between these two components, and with a number of recent works that found the same feature in other clusters \citep[e.g.,][and others; for a brief review of the state-of-the-art literature about BCG/cool core offsets, see \citealp{Pasini_2019}]{Sanderson_2009, Haarsma_2010, Hudson_2010, Rossetti_2016}. We will return on this in Sec. \ref{sec:cavities}.


A two-dimensional temperature map is often used in order to further investigate on the cluster structure and its thermodynamical state, but the small number of photons prevents us from producing such map. We also estimated the \textit{softness ratio} as (S-H)/(S+H), where S and H are the number of counts in the soft (0.5-2 keV) and hard (2-7 keV) band, respectively. However, the statistics are still too poor and errors are too large to draw any conclusion from such analysis.

\subsection{H$\alpha$ analysis}
\label{sec:halpha}

\begin{figure}
	\centering
	\includegraphics[height=24.5em, width=25.5em]{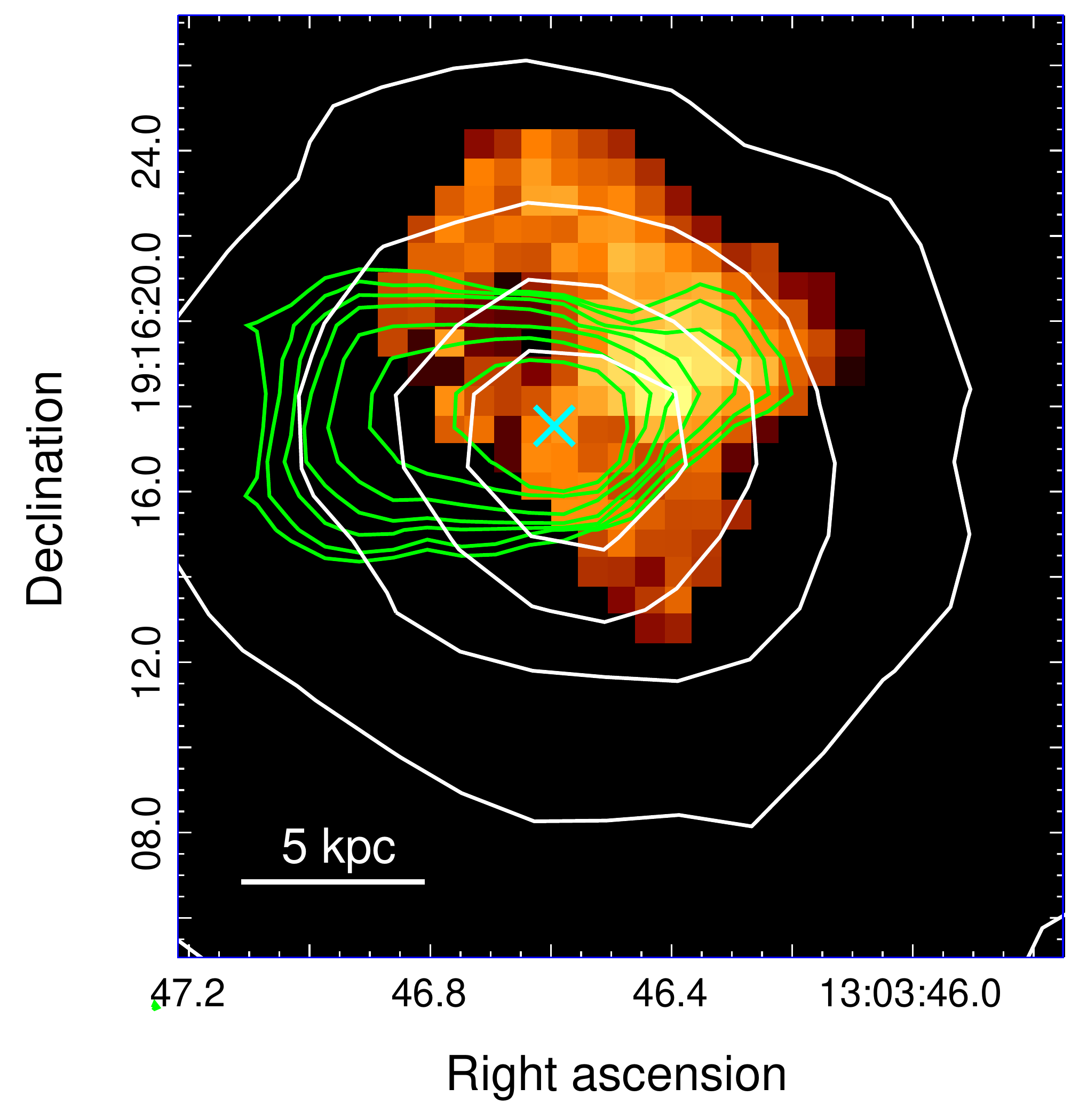}
	\caption{1.4 GHz radio (green) contours overlaid on the \textit{VIMOS} H$\alpha$ image \citep{Hamer_2016}. The mean seeing is 1.21$"$, and the map units are in 10$^{-16}$ erg s$^{-1}$ cm$^{-2} $ \AA. The white contours are the optical isophotes from SDSS. The cyan cross represents the radio galaxy centre, coincident with the BCG core.} 
	\label{fig:halpha}
\end{figure}

The presence of optical line emitting nebulae in galaxy clusters is linked to the thermodynamical conditions of the cluster core; observational studies \citep[e.g.,][]{Cavagnolo_2008, McNamara_2016} have argued that such warm structures are only found if the central ($\sim$ 10 kpc) entropy falls below 30 keV cm$^2$ or, alternatively, when t$_{\text{cool}}$/t$_{ff}$ $<$ 10-20 \citep{Voit_2015}, where t$_{ff}$ = $\sqrt{2R^3/GM}$ is the \textit{freefall} time. The estimated entropy within the central bin of our spectral analysis ($r < 23.5$ kpc,  Table \ref{tab:depro}) is $\sim$ 19.5 keV cm$^2$, thus satisfying the criterion for the presence of such nebulae in A1668.
\citet{Hamer_2016} presented \textit{VIMOS} observations of the H$\alpha$ line emission of a sample of 73 BCGs, including A1668, whose image is shown in Fig. \ref{fig:halpha}.

The H$\alpha$ structure presents a rather compact shape, extending for $\sim$ 9$"$ ($\sim$ 10.6 kpc). Its total luminosity is L$_{\text{H}\alpha}$ = 3.85 $\pm$ 0.30 $\cdot$ 10$^{40}$ erg s$^{-1}$\footnote{This estimate differs from \citet{Pulido_2018} since it is not extinction-corrected and is obtained from IFU observations.}, and \citet{Hamer_2016} classified it as a quiescent object, showing a simple, centrally-concentrated morphology\footnote{The morphology looks slightly different when compared to \citet{Edwards_2009}, whose IFU image shows a somehow better 'resolved' shape.}. To better visualize the interplay of the three components, in Fig. \ref{fig:multi} we also overlay the H$\alpha$ contours on the X-ray 0.5-2 keV image.
\\ \indent
The line emission lies entirely within the BCG, but although it is also within the cool core as defined from the cooling time profile, it only overlaps one end of the bright X-ray ridge. This highlights a difference with A2495, in which the H$\alpha$ structure connects the galaxy with the X-ray peak, and with other systems \citep[e.g.,][]{Bayer-Kim_2002, Hamer_2012, Hamer_2016}, in which the line emission seems to be mostly associated to the cooling ICM, rather than to the BCG. 
\\ \indent
A significant offset of $\sim$ 6.5$"$ ($\sim$ 7.6 kpc) is present between the H$\alpha$ ($RA$=13$^{\text{h}}$03$^{\text{m}}$46.4$^{\text{s}}$,
$DEC$=+19$\degree$16$^{\text{m}}$18.1$^{\text{s}}$) and the X-ray peaks. The same feature, albeit smaller, was found in A2495 by \citet{Pasini_2019}; offsets between these two peaks were also detected in A1795 \citep{Crawford_2005} and in a number of systems by \citet{Hamer_2016}. We will return on this in Sec. \ref{sec:cavities}. 


As described in more details in \citet{Hamer_2012}, the ratio of the H$\alpha$ plume extent and the total velocity gradient of the warm gas provides an estimate of the projected offset timescale. \citet{Pasini_2019} showed that, for A2495, such timescale is comparable to the age difference between two putative cavity pairs, suggesting that, for systems where the BCG oscillates back and forth through the cooling region, this measure could be a good indicator for the AGN cycle intermittency. In A1668, the H$\alpha$ structure extends for D'$\sim$ 7.7$"$ from the BCG centre, corresponding to $\sim$ 9 kpc, and shows a very smooth velocity gradient from East to West. We measured a velocity difference of $\sim$ +500 km s$^{-1}$ between the gas at the BCG centre and that at the tail of the H$\alpha$ structure. This indicates a projected timescale of T'= D'/V $\sim$ 18 Myr.  In order to correct for the projection effects we assumed a most likely inclination of $\sim$ 60 $\degree$, with a range of 30-75 $\degree$ \citep{Hamer_2012}; the corrected timescale can thus be estimated with T = T' $\times$ cos(i)/sen(i), with i being the inclination. We obtained in this way an offset timescale of $\sim$ 10.4 Myr, with an upper and lower limit of $\sim$ 31.2 and $\sim$ 4.8 Myr, respectively; this is consistent with the value of $\sim$ 13 Myr estimated for A2495, suggesting that the cold gas dynamics in the two systems are similar, and that the timescale of AGN feedback intermittency is comparable.

One can also estimate the mass of the warm gas by assuming that it is optically thin:

\begin{equation}
M_{\text{H$\alpha$}} \simeq L_{\text{H$\alpha$}} \dfrac{ 4\mu m_p}{n_{\text{H$\alpha$}}\epsilon_{\text{H$\alpha$}}}
\end{equation}

where $\epsilon_{\text{H}\alpha}  \sim 3.3 \cdot 10^{-25}$ erg cm$^3$
s$^{-1}$ is the H$\alpha$ line emissivity, while $n_{\text{H$\alpha$}}$ is
obtained assuming pressure equilibrium with the local ICM:

\begin{equation}
n_{\text{H$\alpha$}}T_{\text{H$\alpha$}} \simeq n_{\text{ICM}}T_{\text{ICM}}
\label{eq:halphadens}
\end{equation}

where $T_{\text{H$\alpha$}} \sim 10^4$ K, $T_{\text{ICM}}$ is the first value reported in Tab. \ref{tab:proje} ($\sim$ 1.74 keV, since the H$\alpha$ structure is located within the first spectral bin), while $n_{\text{ICM}} \sim 1.83n_e$, where $n_e$ is the electronic density reported in the first row of Tab. \ref{tab:depro}. In this way, we obtain $M_{\text{H$\alpha$}} \sim$ (2.4 $\pm$ 0.2) $\cdot$ 10$^6$ M$_\odot$.

The [SII]$_\lambda 6716$/[SII]$_\lambda 6731$ line ratio provides an independent estimate of the density of the ionised gas and is measured in \citealt{Hamer_2016} (Appendix F) as 1.21$\pm$0.2. Assuming case B reionisation and a temperature of 10$^4$\,K, this gives an electron density of $n_{e,\text{H}\alpha}$\,=350\,$\pm$\,270\,cm$^{-3}$ which corresponds to a total density of $n_{\text{H}\alpha}$=640\,$\pm$\,495\,cm$^{-3}$. These values are comparable, within the (large) uncertainties, with the value of $n_{\text{H}\alpha}$=100\,$\pm$\,$^{8} _{3}$\,cm$^{-3}$ derived from Eq. \ref{eq:halphadens}. It is important to consider the impact of the assumed ionised gas temperature ($T_{\text{H}\alpha}$) on the two measurements though. A lower $T_{\text{H}\alpha}$ would result in a higher $n_{e,\text{H}\alpha}$ from Eq. \ref{eq:halphadens} but a lower $n_{e,\text{H}\alpha}$ from the [SII]$_\lambda 6716$/[SII]$_\lambda 6731$ line ratio measurement, while a higher $T_{\text{H}\alpha}$ would have the opposite effect on both measurements.  The VIMOS data from \citep{Hamer_2016} are not sensitive enough to provide a reliable estimate of $T_{\text{H}\alpha}$ for Abell 1668, but deep observations of other objects have found upper limits to $T_{\text{H}\alpha}$ that are much lower than expected (e.g. $T_{\text{H}\alpha}<$\,5685\,K in the Centaurus cluster, \citealt{Hamer_2019}). Assuming $T_{\text{H}\alpha}$\,=\,5000\,K, we find $n_e$\,=\,230\,$\pm$180\,cm$^{-3}$ and $n_{\text{H}\alpha}$=420\,$\pm$\,330\,cm$^{-3}$ from the [SII] ratio and $n_{\text{H}\alpha}$=200\,$\pm$\,$^{16} _{6}$\,cm$^{-3}$ derived from Eq. \ref{eq:halphadens}, indicating that the two measurements are consistent within the limits of the available data and assumed values.

\subsection{Putative cavities and AGN feedback cycle}
\label{sec:cavities}

\begin{figure}
    \hspace{-0.5cm}
	\centering
	\includegraphics[height=28.5em, width=25.8em]{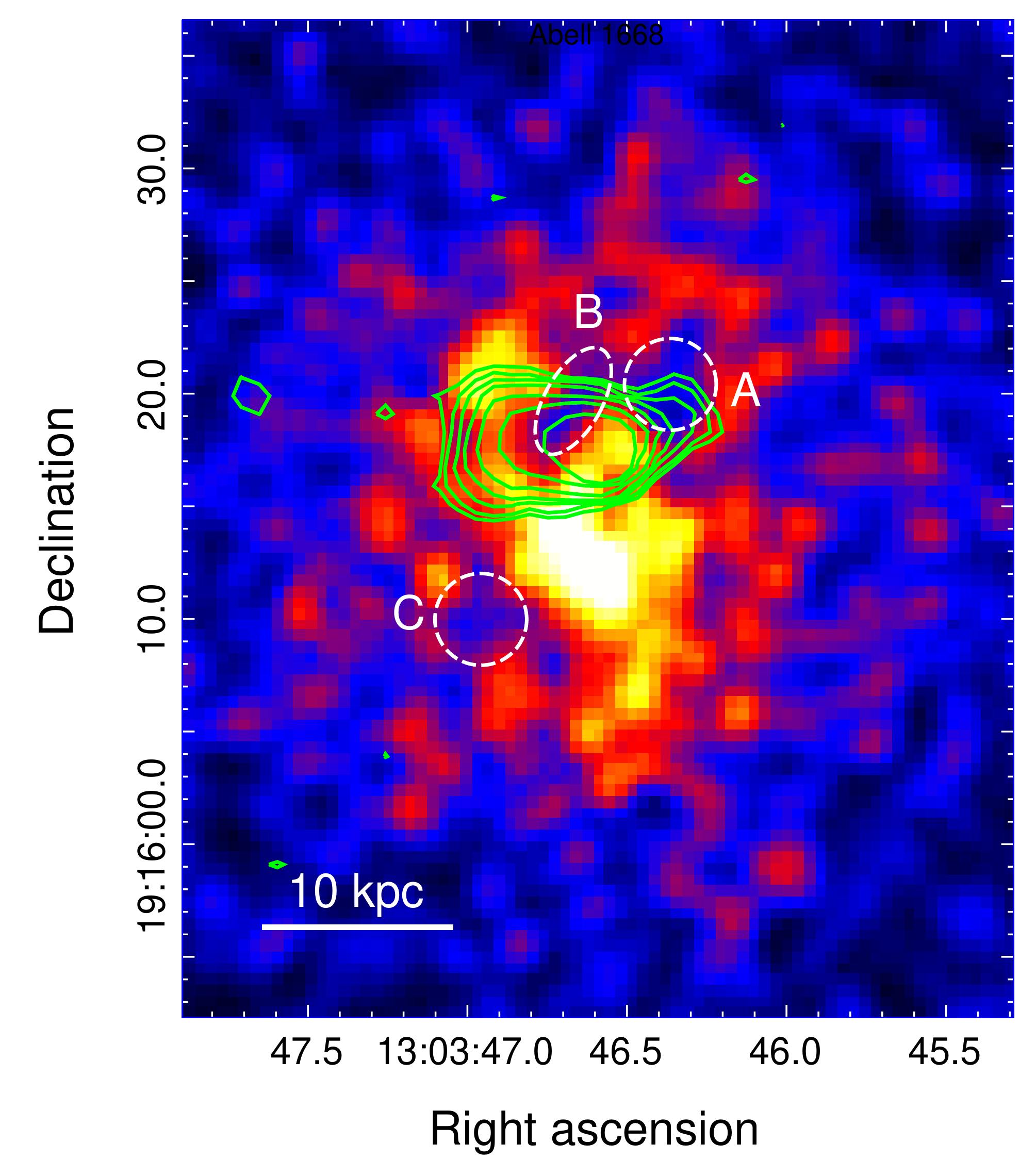}
	\caption{0.5 - 7 keV image showing three surface brightness depressions: A and B lie in the west and east lobes of the radio galaxy, respectively, while C is found at the opposite side of the core. Overlaid are the 1.4 GHz contours.}
	\label{fig:cavity}
\end{figure}

From the 0.5-2 keV X-ray image (see Fig. \ref{fig:cavity}) it is possible to identify a number of ICM surface brightness depressions. The present observation is very shallow ($\sim$ 10 ks), thus the reader shall be warned about the significance of these deficits, that could possibly be artifacts. However, we focused our attention on three of these features (showed in Fig. \ref{fig:cavity}) that, due to their position, could possibly represent real ICM cavities.
One of them (A) lies within the radio galaxy West lobe; another surface brightness depression (B) is detected in the radio galaxy East lobe, while a symmetrical (with respect to the X-ray peak), similar-shaped brightness depression (C), is found at the opposite side of the cluster core, not associated with the radio galaxy. It is noteworthy to mention that cavity A is also coincident with the H$\alpha$ line emission peak.

\begin{figure}
	\centering
	\includegraphics[height=27.5em, width=27.45em]{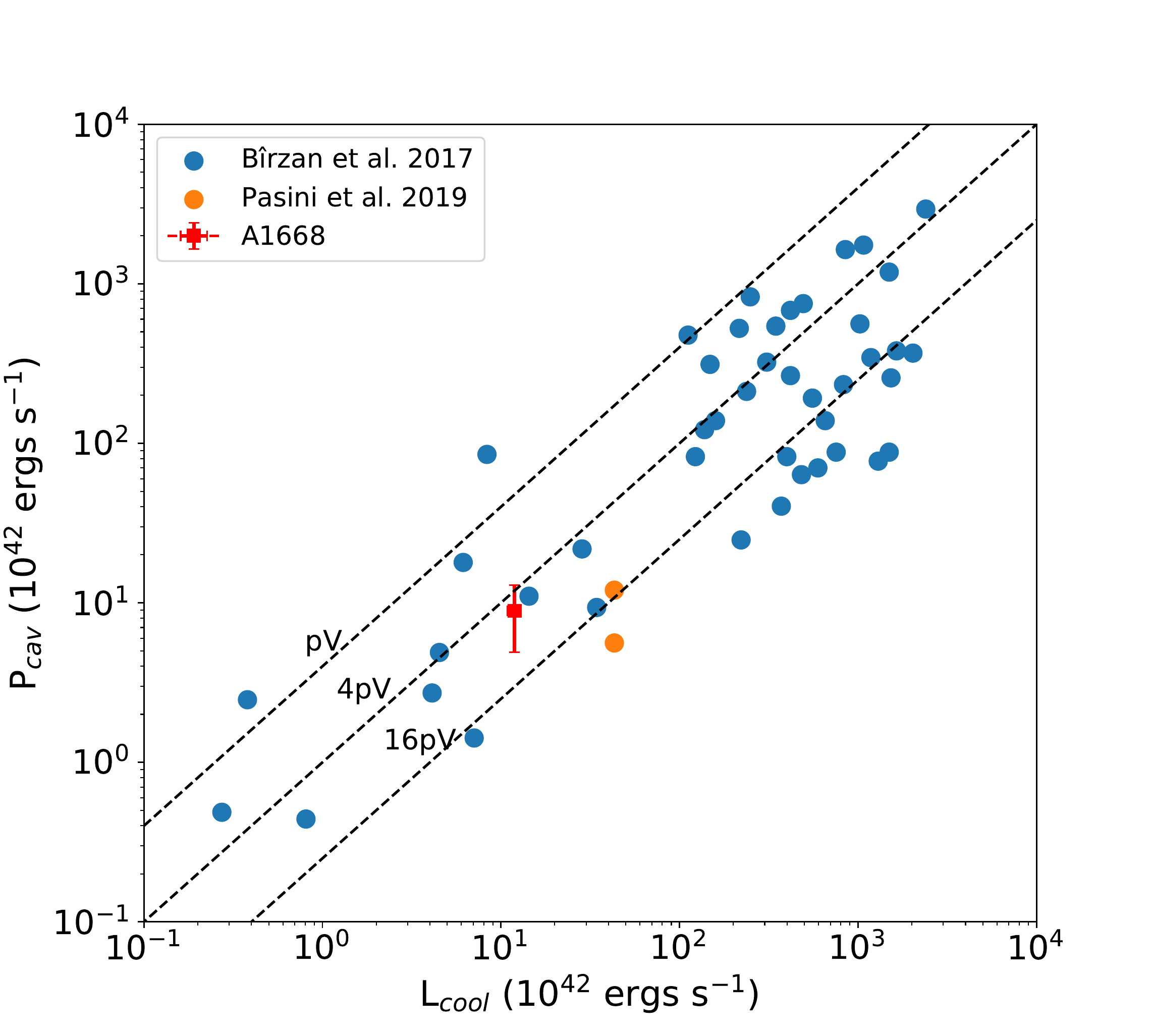}
	\caption{Blue points are the data from \citet{Birzan_2017}, the orange ones are the values for A2495 \citep{Pasini_2019}, while red represents the cavity system detected in A1668. Dashed lines represent, from left to right, P$_{\text{cav}} =$ L$_{\text{cool}}$ assuming $pV$, $4pV$ or $16pV$ as the deposited energy.} 
	\label{fig:confronto}
\end{figure}

To investigate on these features, we estimated their significance as N$_{\text{M}}$-N$_{\text{C}}$/$\sqrt{\text{N}_{\text{M}}+\text{N}_{\text{C}}}$, where N$_\text{M}$ and N$_\text{C}$ are the number of counts in  regions of equal area close to the candidate cavity and within the cavity, respectively. The number of counts is different depending on the size of the elliptical region chosen to cover the putative cavity and, as previously stated, the current observation requires us to be cautious, since the shape and extent of the depression observed 'by eye' can slightly change using different color scales. For this reason, the upper and lower limits for the dimensions of these regions were estimated by varying the axes of the ellipse until reaching a significance of 2$\sigma$ and 3$\sigma$, respectively. The assumed 'true' size of the bubble is the mean between these two limits. Cavity A exhibits a circular shape, with a diameter of 4.8 $\pm$ 0.6 kpc, while cavity B is more elliptical, with a major axis of 5.2 $\pm$ 0.4 kpc and a minor axis of 3.0 $\pm$ 0.3 kpc. On the other hand, the brightness deficit related to candidate cavity C is less enhanced and, in order to reach the desired significances, requires the size to be larger than the observed depression. Therefore, in the following analysis we will only discuss candidate cavities A and B, while cavity C will not be considered. Following the method described in \citet{Birzan_2004}, we determined the cavity power:

\begin{equation}
P_{\text{cav}} = \dfrac{E_{\text{cav}}}{t_{\text{cav}}} = \dfrac{4pV}{t_{\text{cav}}}
\end{equation}
\label{pcav}

where $t_{\text{cav}}$ is the age of the cavity, calculated as $t_{\text{cav}} = R/c_s$, with \textit{R} being the cavity distance from the BCG center, $p$ is the pressure at the distance of the cavity, $V$ is the cavity volume and $c_s$ is the sound velocity. The volume was estimated assuming an oblate elipsoidal shape for both cavities, while for pressure and temperature we assumed the values listed in Table \ref{tab:depro} corresponding to the annular bin the cavities lie in. 
\\ \indent
We obtained the same age for both cavities: t$_{\text{cav}}$ = 5.2 $\pm 0.7$ Myr. This leads to P$_{\text{cav, A}}$ = 5.1 $\pm 2.6 \ \cdot$ 10$^{42}$ erg s$^{-1}$ for cavity A, and P$_{\text{cav, B}}$ = 3.8 $\pm 1.4 \ \cdot$ 10$^{42}$ erg s$^{-1}$ for cavity B. The age is consistent with the offset timescale estimated in Sec. \ref{sec:halpha}.
Finally, we compared the estimated values of P$_{\text{cav}}$ (in the hypothesis that the cavities are real) and L$_{\text{cool}}$ with the typical distribution observed for cool core clusters \citep{Birzan_2017}. The result is presented in Fig. \ref{fig:confronto}.

The L$_{\text{cool}}$-P$_{\text{cav}}$ relationship in A1668 is consistent within the scatter of the expected distribution, despite being on its lower edge. The detected offsets, therefore, do not seem to affect the feeding-feedback cycle, that is still maintained. The same result was found in A2495, where the two cavity systems (i.e., the AGN) have enough energy to balance the radiative losses within the cooling region. We then argue that small offsets are not able to break the AGN feeding-feedback cycle.

\subsection{Offsets, cooling and H$\alpha$ emission: are sloshing and AGN activity shaping the core of A1668?}
\label{sec:discussion}

As described in the previous sections, the core of A1668 contains a complex set of structures whose origin is not immediately clear. It is worth reiterating that the available \textit{Chandra} data is only a snapshot, providing somewhat limited information on the ICM. It should also be remembered that all of the structures we observe are contained within the central $\sim$ 20 kpc, inside the stellar body of the BCG.

The offsets between the peaks of the radio, X-ray and H$\alpha$ emission raise the question of how these components came to be separated. The peak of emission from the hot ICM lies not in the BCG nucleus, but $\sim$6~kpc to the south. The radio emission is reasonably evenly distributed Eastwards and Westwards with respect to the nucleus, but the H$\alpha$ peaks in the region of the west radio jet/lobe, extending around the western half of the radio structure and overlapping the optical centroid of the BCG.

We suggest a qualitative scenario which might explain the relative morphologies of the different components. At some point in the past, A1668 may have been a relaxed cluster with a cool core. In that core, centred on the BCG, gas had begun to cool and condense out of the ICM, forming the kind of filamentary H$\alpha$ nebula observed in other cool core cluster. The small velocity gradient in the H$\alpha$ emission is consistent with such an origin. At that stage, A1668 underwent a minor merger, which caused the core to begin sloshing, oscillating around the centre of the cluster gravitational potential. Sloshing motions in the plane of the sky are typically visible as a spiral pattern in the ICM, but if the plane of motion is aligned along the line of sight, the motions produce pairs of nested cold fronts, and the cool ICM gas drawn out from the core can appear as a tail to one side of the BCG. Our \textit{Chandra} observation is too short for fronts to be visible, but we do see the tail: the ridge structure.

At this stage, cooling (and H$\alpha$ emission) would still have been centred in the core of the BCG. About 5 Myr ago ($\sim$ cavity age), sufficient cooled material reached the central SMBH to trigger an outburst. This produced the radio jets/lobes we observe, and as these expanded they pushed aside the pre-existing H$\alpha$ filaments and hotter ICM gas. This produced a correlated H$\alpha$/radio morphology, with much of the H$\alpha$ wrapped around the west jet/lobe. It also disrupted the centre of the cool core, reducing the X-ray surface brightness as a large part of the volume in the core of the BCG was filled by the radio lobes, producing the apparent cavities in the ICM. This brings us to the current situation, where the brightest X-ray emission is in the tail to the south of the BCG nucleus.

The expansion timescale of the radio lobes is only a few 10$^6$~yr. This is very short compared to typical sloshing timescales. The hot ICM oscillates with sloshing
timescale given by $t_{\rm slosh} = 2\pi/\omega_{\rm BV}$ where $\omega_{\rm BV}= \Omega_K
\left(\frac{1}{\gamma}\frac{d\ln S}{d\ln r}\right )^{1/2}$. Here $d$ls$S$/$d$ln$r$ is the logarithmic entropy gradient, $\Omega_K$=$\sqrt{GM/r^3}$ and $\gamma$=5/3 for the ionized ICM plasma. We also know that the free-fall time in the cluster is $t_{\rm ff}=\sqrt{2r/g}$. Our \textit{Chandra} data are not sufficient to accurately model the mass profile, but we know that the stellar velocity dispersion, $\sigma_*$, in the inner regions of the BCG will follow the gravitational potential, so that $t_{\rm ff}\simeq r/\sigma_*$ \citep{Voit_2015}. We can therefore approximate $\omega_{\rm BV}$ as $\frac{1}{t_{ff}}\sqrt{\frac{6}{5}\frac{d\ln S}{d\ln r}}$.

Our data do not allow the calculation of the entropy profile on scales $r\lesssim 25$ kpc (Fig. \ref{fig:entropy}), thus we use the average slope $d{\rm ln}S/d{\rm ln}r=0.67$ given by \citet{Hogan_2017} (see also \citealt{Panagoulia_2014a}). 
This returns $t_{\rm slosh}\sim 7 t_{\rm ff}$. We do not know the scale of the sloshing, but it must be greater than the length of the X-ray tail ($\sim$16 kpc). Based on the measured $\sigma_*$=226$\pm$7~km~s$^{-1}$ \citep{Pulido_2018}, at 16 kpc $t_{\rm ff}$=70 Myr, and thus $t_{\rm slosh}$ may be as much as $\sim$ 490 Myr. As expected this is considerably longer than the AGN expansion timescale, confirming that if sloshing is occurring, it cannot yet have affected the structure of the radio lobes.

As argued by \citet{Olivaresetal19}, the fact that filamentary nebulae in cool core clusters generally lack a significant velocity gradient indicates that the cool H$\alpha$ or CO-emitting gas they contain is at least partially tied to the surrounding ICM. It is unclear how the different phases are connected, but it has been suggested that the denser material may be enveloped by many diffuse layers of warmer gas \citep{Lietal18}, or threaded through by magnetic fields \citep{McCourtetal15}, either of which could increase drag forces.
 We would thus expect the H$\alpha$ emission to trace the regions in which gas has most recently cooled from the ICM, after modification by the expanding radio lobes. If the inflation of cavities has disrupted the cooling region, we might expect the locus of any future cooling to be at the new X-ray peak of surface brightness, source of the BCG nucleus. However, the lack of H$\alpha$ emission at that location suggests that cooling there is slower than it was near the BCG core, and that no reservoir of cooled material has yet built up at that location.

This scenario is of course speculative, given the constraints available from the data. Several aspects are uncertain. All, part or none of the H$\alpha$ emitting gas might have formed as a result of the AGN outburst, with the expanding lobes triggering condensation \citep[e.g.,][]{Qiu_2020}. If the cluster is sloshing, we cannot know the scale or alignment of the motion without deeper data. However, our scenario explains several basic facts: the radio and H$\alpha$ emission are correlated because the radio has at least partly determined the morphology of the H$\alpha$-emitting gas. The X-ray offset is the result of sloshing, which has not affected the radio sources or H$\alpha$ because the radio source expansion timescale is short compared to the sloshing timescale. The BCG is no longer the centre of ICM cooling because the AGN has pushed aside the dense gas which fuelled the outburst.

The scenario also makes testable predictions. If the cluster is sloshing, we should expect deeper \textit{Chandra} data to reveal nested cold fronts, and the X-ray ridge should contain relatively cool, high abundance gas. Deeper imaging should also allow us to more accurately measure the morphology of any cavities, which should be correlated with the rado jets and lobes. Higher resolution radio data may be required to make this comparison. Lastly, we might expect higher resolution H$\alpha$ imaging to reveal complex structure in the cooled material, consistent with a filamentary nebula disturbed by an AGN outburst.

\subsection{Alternative explanations for the origin of the H$\alpha$ emission and spatial offsets}
\label{sec:discussion2}

An alternative hypothesis that could explain the observed displacement of the
H$\alpha$ emission is cooling \textit{in situ}, perhaps stimulated by the same AGN outburst 
which originated the X-ray cavities. 
Inhomogeneous cooling scenarios in clusters have been the object of a
long, lively debate (see early reviews by \citealt{Fabian_1991, Fabian_1994}, and
references therein). Nowadays, many lines of
evidence suggest that hot gas cools at a (mean) low
rate and in a spatially distributed fashion, when the ISM/ICM
conditions are appropriate (see \citealt{Hogan_2017, Pulido_2018, Lakhchaura_2018}
for a quantitative discussion).
The primary trigger of this localized cooling (that is, the origin of
thermally unstable perturbations) might be turbulence \citep[e.g. Chaotic Cold Accretion, CCA, see][]{Gaspari_2012, Gaspari_2013, Voit_2015}, lifting of low entropy gas by X-ray cavities \citep{Revaz_2008, Brighenti_2015} or the sloshing itself.

CCA implies the trigger of thermal instabilities, that is favoured by a central ($\leq$ 10 kpc)
cooling time $t_{\rm cool} \sim 1$ Gyr \citep{Hogan_2017b, Pulido_2018}, or by 
$t_{\rm cool}/t_{ff} \leq 10-20$ \citep{Voit_2015}. 
It is easier to ensue at the position of the X-ray peak, but it can be triggered 
wherever these conditions are respected. Therefore, the displacement observed for
the H$\alpha$ gas could be the result of CCA detached from the emission peak. CCA
at the current position of the peak could still be happening, but it could have not built 
yet enough material to be detected in H$\alpha$.
However, given the available X-ray observation, we are not in a position to accurately
estimate the cooling time in the central region of the cluster.

We can explore in some more detail the
scenario where the warm gas derives from a cool component of the ICM, originally
located close to the nucleus of the BCG, uplifted by
the cavities and then cooled to $10^4$ K.
Following Archimedes' principle, cavities can lift an amount of gas equal to their
displacement, though simulations suggest that the maximum
amount is only $\sim$50\% of this value \citep{Pope_2010b}. This
corresponds to M$_{\text{uplift}}$ = 9 $\cdot$ 10$^6$ M$_{\odot}$ for
cavity A, and M$_{\text{uplift}}$ = 3 $\cdot$ 10$^6$ M$_{\odot}$ for
cavity B. The mass of the H$\alpha$ plume is lower
($M_{\text{H$\alpha$}} \sim$ 2.4 $\cdot$ 10$^6$ M$_\odot$, see Sec.
\ref{sec:halpha}). However, given the state-of-the-art correlation
between L$_{\text{H}\alpha}$ and molecular gas mass M$_{\text{mol}}$
\citep[see e.g.,][]{Edge_2001, Salome-Combes_2003, Pulido_2018}, that
is usually found to be co-spatial with H$\alpha$, we would expect to
have M$_{\text{mol}}$ $\sim$ 10$^9$ M$_{\odot}$ (lower than the upper
limit $1.5\times 10^9$ M$_\odot$ quoted by \citealt{Salome-Combes_2003}).
The total amount of gas would therefore be too large for the cavities to uplift and this,
along with the radio/H$\alpha$ morphology, would imply the need for an
earlier cycle of AGN jet activity if uplift is responsible.

The observed H$\alpha$ line emission could also be the remnant of the ISM of a gas-rich galaxy which merged
with the BCG. To test this hypothesis, we examined
SDSS and DSS optical images and catalogs in order to check whether a
member galaxy could be interacting with the BCG. However, the closest
system lies more than 40 kpc away from the BCG, not showing any hint
of interplay. Therefore, the merging hypothesis looks unlikely with
the current data.

Finally, the warm gas could originate from the stellar mass loss in the
BCG \citep{Mathews_1990, Li_2019}. With a total B-band
luminosity $L_B \sim 1.3\times 10^{11}$ $L_{B,\odot}$ \citep{Makarov_2014} 
and a stellar mass to ratio for an old population $(M/L_B)
\sim 7$ \citep[e.g.,][]{Maraston_2005}, the expected mass loss rate is $\dot M_*
\sim 1.35$ M$_\odot$/yr \citep{Mathews_1989}. Thus, the observed amount of
emission line gas can be accumulated in less than 2 Myr. However, the
displacement with respect to the BCG center and its filamentary and
disturbed distribution \citep{Edwards_2009, Hamer_2016} are
not easily accounted for by this scenario.

\section{Conclusions}

We performed a multi-wavelength analysis of the cool core cluster A1668, by means of new radio (EVLA) and X-ray (\textit{Chandra}) observations and of H$\alpha$ line emission data from \citet{Hamer_2016}. The results can be summarized as follows:

\begin{itemize}
	
	\item The radio analysis at 1.4 (L$_{\text{1.4 GHz}} \sim 6 \cdot 10^{23}$ W Hz$^{-1}$) and 5 GHz (L$_{\text{5 GHz}} \sim 2 \cdot 10^{23}$ W Hz$^{-1}$) shows a small ($\sim$ 11 -14 kpc) and elongated FRI radio galaxy, with no hints of larger scale emission at these frequencies. The mean spectral index is $\alpha = 0.99 \pm 0.06$, consistent with the usual values found in BCGs.
	
	\item The X-ray analysis confirms the classification of A1668 as a cool core cluster, with a cooling radius of $\sim$ 40 kpc  inside which we estimate a bolometric luminosity  $L_{\rm cool }\sim$ 1.9 $\cdot$ 10$^{43}$ erg s$^{-1}$.
	
	\item The multi-wavelength analysis reveals two spatial offsets, with the first of $\sim$ 6 kpc being between the BCG nucleus and the X-ray peak, while the second of $\sim$ 7.6 kpc between the H$\alpha$ and the X-ray peaks. This is similar to what was found in another similar cluster, A2495, with two offsets of 6 and 4 kpc, respectively \citep{Pasini_2019}. The compact H$\alpha$ emission structure extends for $\sim$ 11 kpc and is mostly co-spatial with the BCG, unlike A2495, where the line emission seems to be linked to the cluster cool core, rather than to the central galaxy. 
	
	\item We identify three X-ray surface brightness depressions, one of them (A) coincident with the west radio lobe and with the H$\alpha$ peak, another one (B) lying within the east radio lobe, while the third one (C) being more uncertain. For the system of cavities A and B we determine an age of $\sim$ 5.2 Myr. The L$_{\text{cool}}$-P$_{\text{cav}}$ estimates for A1668 are in agreement with the relationship observed for other systems \citep[e.g.,][]{Birzan_2017}, suggesting that the detected offsets are not able to break the AGN feeding-feedback cycle.
	
    \item Finally, we discuss possible explanations for the multiphase gas and for the displacements observed in the core of A1668. We propose that, initially, all the components were spatially coincident in the cluster cool core. Sloshing was likely triggered by a minor merger, causing some of the cool gas around the BCG to be drawn out into a tail that we now observe as an X-ray ridge structure. On the other hand, the densest, most rapidly cooling gas, still in and around the BCG core, condensed out to form the H$\alpha$ nebula. About 5 million years ago, the condensed material fuelled the central SMBH, triggering the outburst that produced the observed radio jets/lobes. The expansion of the lobes finally pushed aside the H$\alpha$ nebula and the hot ICM, disrupting the cool core centre. Alternative explanations for the misplacement of the H$\alpha$ emission include cooling \textit{in situ} through thermal instabilities, uplift from the cavities, reminiscence from a past merger with a gas-rich galaxy, or stellar mass loss from the BCG, although the last three look unlikely (see Sec. \ref{sec:discussion2}).

\end{itemize}

\acknowledgments
We thank the referee for thoughtful comments and suggestions, which have significantly improved the presentation of our results. TP is supported by the BMBF Verbundforschung under grant number 50OR1906. E.O'S. ackonwledges support for this work from the National Aeronautics and Space Administration through Chandra Award Number G07-18162X issued by the Chandra X-ray Center, which is operated by the Smithsonian Astrophysical Observatory for and on behalf of the National Aeronautics Space Administration under contract NAS8-03060. Based on observations made with ESO Telescopes at the La Silla or Paranal Observatories under programme ID 080.A-0224 and 082.B-0671. The National Radio Astronomy Observatory is a facility of the National Science Foundation operated under cooperative agreement by Associated Universities, Inc.

\bibliographystyle{aasjournal} 
\bibliography{bibliography.bib}

\end{document}